\documentclass[
floatfix,
aps,
amsmath,
nofootinbib,
onecolumn,
10pt,
]{revtex4}

\usepackage{graphicx}
\usepackage{bm}
\usepackage{rotating}
\usepackage{array}

\begin{document}

\title{A Gravitational Wave Background from Primordial Black Hole Lattices in Matter Dominated Era}

\author{
Ke Wang \footnote{wangke@itp.ac.cn}
}

\affiliation{
National Astronomical Observatories, \\Chinese Academy of Sciences,  20A Datun Road, Beijing 100012, China\\
}

\date{\today}

\begin{abstract}
We use the wide-used \textsf{Einstein Toolkit} to solve the Einstein constraints and then simulate the expansion of primordial black hole lattices (PBHLs) with different value of $f_{\mathrm{PBH}}$ and $m_{\mathrm{PBH}}$. We find that $f_{\mathrm{PBH}}$ plays an important role during the evolution of PBHLs. Since the motion of primordial black holes (PBHs) caused by the expansion of PBHLs occurs at speeds close to that of light, we expect the emission of gravitational waves (GWs) during the expansion of PBHLs. We use both analytical estimates and numerical simulations to cross check the production of GWs in expanding PBHLs.

\end{abstract}


\maketitle


\section{Introduction}
Observations of gravitational waves (GWs) by LIGO/VIRGO \cite{Abbott:2016blz,Abbott:2016nmj,Abbott:2017vtc,Abbott:2017oio,TheLIGOScientific:2017qsa} have proved there should be a gravitational wave background (GWB) produced by the black hole and neutron star binaries' coalescence.
Meanwhile, people also believe there is a primordial inflationary GWB from tensor perturbations generated by quantum fluctuations during inflation even though Planck didn't measure it \cite{Ade:2015xua}. Besides that, there are many other theories (or sources) which can produce GWBs, such as reheating after inflation \cite{GarciaBellido:2007af}, thermal phase transitions from the decays of cosmic strings \cite{Maggiore:1999vm}.
Here we expect that there is a GWB from primordial black hole lattices (PBHLs) in matter dominated era.

To what extend can the configuration of primordial black holes (PBHs) just after the epoch of matter-radiation equality be considered as a PBHL? As we known, a pair of PBHs would decouple from the expansion of the Universe and form a gravitationally bound system if the average energy density of PBHs over the volume $R^3$ is larger than the total background cosmic energy density $\rho_t$, that is, if $m_{\mathrm{PBH}}R^{-3}>\rho_t$.
Since $\rho_t\approx m_{\mathrm{PBH}}\bar{R}^{-3}f_{\mathrm{PBH}}^{-1}$, we know that most of PBHs are gravitationally free on average in matter dominated era.
On the other hand, due to the cosmological principle that the Universe is homogeneous and isotropic on large scales, we can say that the PBHL is the most reasonable configuration of PBHs with a monochromatic mass distribution on our hands.

In fact, there is a cosmological model, black hole lattices (BHLs) \cite{Lindquist,Yoo:2012jz,Yoo:2013yea,Bentivegna:2013jta}, to investigate the so-called ``backreaction'' that the local inhomogeneities can affect the global expansion of the Universe. For review of BHLs, see \cite{Bentivegna:2018koh}.
It's worth pointing out that the PBHLs doesn't serve as a cosmological model like BHLs does but is an early epoch of the Universe. In practice, however, we  do apply the same technologies developed by \cite{Yoo:2012jz,Yoo:2013yea,Bentivegna:2013jta} when they evolve the BHLs in numerical relativity to the PHBLs.

Here we will turn to the wide-used \textsf{Einstein Toolkit} \cite{Loffler:2011ay} to solve the Einstein constraints and simulate the spacetime and hydrodynamical evolution. More precisely: the thorn \textsf{CT\_MultiLevel} (and \textsf{CT\_Analytic}) \cite{Bentivegna:2013xna} gets the initial data of PBHLs; the thorn \textsf{McLachlan} \cite{Brown:2008sb,Reisswig:2010cd,code} evolves spacetime using the Baumgarte-Shapiro-Shibata-Nakamura (BSSN) formalism \cite{Baumgarte:1998te,Shibata:1995we,Alcubierre:2000xu}; the thorn \textsf{GRHydro} evolves the hydrodynamical system \cite{Moesta:2013dna,Baiotti:2004wn,Hawke:2005zw}.

This paper is organized as follows. In Subsec.~\ref{constraints}, we give the Einstein constraints of PBHLs with dust. In Subsec.~\ref{solution}, we solve the Einstein constraints of PBHLs with different value of $f_{\mathrm{PBH}}$ and $m_{\mathrm{PBH}}$. In Subsec.~\ref{expansion}, we show the expansion of PBHLs with different value of $f_{\mathrm{PBH}}$ and $m_{\mathrm{PBH}}$. In Subsec.~\ref{production}, we use both analytical estimates and numerical simulations to derive the production of GWB in PBHLs. At last, a brief summary and discussion are included in Sec.~\ref{summary}.

In this paper, we adopt the following conventions: Greek indices run in \{0, 1, 2, 3\}, Latin indices run in \{1, 2, 3\} and repeated indices implies summation and we are in a geometric unit system with $G=c=M_{\odot}=1$.

\section{Initial Data of Primordial Black Hole Lattices}
To mock an infinite 3-dimensional lattice in our simulations, we impose the periodic boundary conditions on a cubic with $x^i$ in $[-5,5]$. A PBH is located at the center of cubic and surrounded by flat-distributed dust with energy density $\rho_m$. It's worth noting that our simulation is not done in Schwarzschild coordinates but in its isotropic coordinates. For simplicity, we just add the dust outside the PBH horizon $r_h=0.5m_{\mathrm{PBH}}$ so that the inner boundary conditions of PBHLs can be set as that of a single puncture \cite{Brandt:1997tf}. And there are no outer boundary conditions for PBHLs.
Although the set up of initial data is usually a non-trivial task, here we can solve PBHLs' coupled system of one non-linear and three linear elliptic partial differential equations thanks to \cite{Yoo:2012jz,Bentivegna:2013jta,Bentivegna:2013xna}.
\subsection{Einstein Constraints of Primordial Black Hole Lattices}
\label{constraints}
The initial data of PBHLs must satisfy their Hamiltonian constraint and momentum constraint
\begin{eqnarray}
  \nonumber
  R+K^2-K_{ij}K^{ij}&=& 16\pi E, \\
  D_jK^j_i-D_iK&=& 8\pi p_i,
\end{eqnarray}
where the 3-metric $\gamma_{ij}$ is the intrinsic metric, $K_{ij}$ is the extrinsic curvature, $R$ is 3-Ricci scalar, $D_i$ is the covariant derivative associated with $\gamma_{ij}$, $E$ is the matter energy density and $p_i$ is the matter momentum density as measured by the Eulerian observer. After conformal decomposition of  $\gamma_{ij}$ and $K_{ij}$ with the conformal factor $\Psi$
\begin{eqnarray}
  \nonumber
  K_{ij} &=& A_{ij}+\frac{1}{3}K\gamma_{ij},\\
  \nonumber
  \gamma_{ij} &=& \Psi^{4}\tilde{\gamma}_{ij},\\
  A_{ij} &=& \Psi^{-10}\hat{A}_{ij},
\end{eqnarray}
we can get the conformal Einstein constraints
\begin{eqnarray}
  \nonumber
  \tilde{D}_i\tilde{D}^i \Psi-\frac{1}{8}\tilde{R}\Psi+\frac{1}{8}\hat{A}_{ij}\hat{A}^{ij}\Psi^{-7} +2\pi E\Psi^{5}-\frac{1}{12}K^2\Psi^5 &=& 0,\\
  \tilde{D}_{j} \hat{A}^{ij}-\frac{2}{3}\Psi^6\tilde{D}^iK &=& 8\pi \Psi^{10}p^i,
\end{eqnarray}
where $\tilde{D}_i$ is the covariant derivative associated with the conformal metric $\tilde{\gamma}_{ij}$. Adopting the conformal transverse traceless method
\begin{eqnarray}
  \nonumber
  \hat{A}^{ij} &=& \tilde{D}^iX^j+\tilde{D}^jX^i-\frac{2}{3}\tilde{D}_kX^k\tilde\gamma^{ij}+\hat{A}^{ij}_{\text{TT}},\\
  \tilde{D}_{j} \tilde{A}^{ij} &=&  \tilde{D}_j\tilde{D}^jX^i+\frac{1}{3}\tilde{D}^i\tilde{D}_jX^j+\tilde{R}^i_jX^j,
\end{eqnarray}
and choosing the free data as following
\begin{eqnarray}
  \nonumber
  \tilde\gamma^{ij} &=&\delta^{ij},\\
  \nonumber
  \hat{A}^{ij}_{\text{TT}} &=&  0,\\
  \nonumber
  E &=& \Gamma^2(\rho_m+P_m)-P_m=\rho_m,\\
  \nonumber
  p^i &=&(E+P_m)v^i_m=0,\\
  K &=& K_cW(r)=K_c\begin{cases}
 0,~~0\leq r\leq l\\
 \left(\frac{(r-l-\sigma)^6}{\sigma^6}-1\right)^6,~~l<r<l+\sigma\\
 1,~~l+\sigma\leq r\leq L,
 \end{cases},
\end{eqnarray}
we can get the conformal transverse traceless form of Einstein constraints for PBHLs
\begin{eqnarray}
  \label{Econstraints}
  \nonumber
  \triangle \psi+\triangle\left(\frac{m_{\mathrm{PBH}}}{2r}W(r)\right)-\frac{1}{12}K^2\Psi^5+\frac{1}{8}\hat{A}_{ij}\hat{A}^{ij}\Psi^{-7} +2\pi\rho_m\Psi^5 &=& 0,\\
  \triangle X^i+\frac{1}{3}\partial^i\partial_jX^j-\frac{2}{3}\Psi^6\partial^iK &=& 0,
\end{eqnarray}
where we have expanded the conformal factor as
\begin{equation}
 \Psi =\psi+\frac{m_{\mathrm{PBH}}}{2r}(1-W(r))
\end{equation}
according to the inner boundary conditions of a single puncture.

\subsection{Solving the Einstein Constraints}
\label{solution}
We use a multigrid approach \cite{Bentivegna:2013xna} to solving the Einstein constraints (\ref{Econstraints}). But we impose the integrability condition as \cite{Yoo:2012jz}
\begin{equation}
 2\pi m_{\mathrm{PBH}}-\frac{1}{12}K_c^2\int W^2\Psi^5+\frac{1}{8}\int\hat{A}_{ij}\hat{A}^{ij}\Psi^{-7} +2\pi\int\rho_m\Psi^5 =0,
\end{equation}
at the end of each relaxation step to determine the negative parameter $K_c$. Finally we perform the inner resetting
\begin{eqnarray}
  \nonumber
  \psi(0) &=&1,\\
  X^i (0)&=&0,
\end{eqnarray}
at the end of every relaxation step.

Tab.~\ref{tab:PBHLsetting} shows the parameters setting of ten PBHLs and Tab.~\ref{tab:PBHLsolution} shows the solution of the Einstein constraints of ten PBHLs when all of $5$ refinement levels cover the whole domain with spacing $1$, $0.5$, $0.25$, $0.125$, and $0.0625$ respectively. Comparing L0 to L7, L8 and L9, we can see that a smaller PBH mass $m_{\mathrm{PBH}}$ gives a shorter initial proper cubic edge $D_{\mathrm{edge}}(\tau=0)$ and leads a lower initial expansion rate $-K_c$. Comparing L0 to L4, L5 and L6, we can see that a higher matter energy density $\rho_m$ has a higher initial expansion rate $-K_c$ but produces a shorter initial proper cubic edge $D_{\mathrm{edge}}(\tau=0)$. Comparing L0 to L1, L2 and L3, we can see that a larger PBH mass $m_{\mathrm{PBH}}$ and a smaller matter energy density $\rho_m$ still produce a larger initial proper cubic edge $D_{\mathrm{edge}}(\tau=0)$ but lead a lower initial expansion rate $-K_c$. That is to say, the initial expansion rate $-K_c$ is more sensitive to the matter energy density $\rho_m$ than the PBH mass $m_{\mathrm{PBH}}$. From the Tab.~\ref{tab:PBHLsolution}, we can also see that a smaller PBH mass $m_{\mathrm{PBH}}$ violates the Hamiltonian constraint more severely.
\begin{table*}[!htp]
\centering
\renewcommand{\arraystretch}{1.5}
\begin{tabular}{cccccccc}
\hline\hline
Lattices&$f_{\mathrm{PBH}}$&$m_{\mathrm{PBH}}$&$\rho_m$&$\rho_t$ &$L$&$\sigma$&$l=\frac{\mathrm{PBH}}{2}$ \\
\hline
L0&$100\%$ &$2$    &$0$       &$0.002$  &$5$  &$3.5$  &$1$ \\
\hline
L1&$75\%$  &$1.5$  &$0.0005$  &$0.002$  &$5$  &$3.75$  &$0.75$ \\
L2&$50\%$  &$1$    &$0.001$   &$0.002$  &$5$  &$4$  &$0.5$\\
L3&$25\%$  &$0.5$  &$0.0015$  &$0.002$  &$5$  &$4.25$  &$0.25$\\
\hline
L4&$80\%$  &$2$  &$0.0005$  &$0.0025$  &$5$  &$3.5$  &$1$ \\
L5&$66.7\%$  &$2$    &$0.001$   &$0.003$  &$5$  &$3.5$  &$1$\\
L6&$62.5\%$  &$2$  &$0.0015$  &$0.0035$  &$5$  &$3.5$  &$1$\\
\hline
L7&$100\%$  &$1.5$  &$0$  &$0.0015$  &$5$  &$3.75$  &$0.75$ \\
L8&$100\%$  &$1$    &$0$   &$0.001$  &$5$  &$4$  &$0.5$\\
L9&$100\%$  &$0.5$  &$0$  &$0.0005$  &$5$  &$4.25$  &$0.25$\\
\hline\hline
\end{tabular}
\caption{The parameters setting of ten PBHLs.}
\label{tab:PBHLsetting}
\end{table*}
\begin{table*}[!htp]
\centering
\renewcommand{\arraystretch}{1.5}
\begin{tabular}{cccccccc}
\hline\hline
Lattices&$f_{\mathrm{PBH}}$&$m_{\mathrm{PBH}}$&$K_c$&$D_{\mathrm{edge}}(\tau=0)$&$H_{\mathrm{max}}$&$H_{\mathrm{min}}$ \\
\hline
L0&$100\%$ &$2$    &$-0.2480$     &$14.16$     &$0.0023$   &$-0.0009$ \\
\hline
L1&$75\%$  &$1.5$  &$-0.3413$     &$12.03$     &$0.0032$   &$-0.0021$ \\
L2&$50\%$  &$1$    &$-0.3864$     &$10.76$     &$0.0168$   &$-0.0114$ \\
L3&$25\%$  &$0.5$  &$-0.4000$     &$10.07$     &$0.2438$   &$-0.2868$ \\
\hline
L4&$80\%$  &$2$    &$-0.3711$     &$12.63$     &$0.0047$   &$-0.0021$ \\
L5&$66.7\%$  &$2$  &$-0.4869$     &$11.21$     &$0.0087$   &$-0.0041$ \\
L6&$62.5\%$  &$2$  &$-0.6200$     &$9.84$      &$0.0178$   &$-0.0060$ \\
\hline
L7&$100\%$  &$1.5$ &$-0.2338$     &$13.12$     &$0.0029$   &$-0.0020$ \\
L8&$100\%$  &$1$   &$-0.2120$     &$12.07$     &$0.0168$   &$-0.0114$ \\
L9&$100\%$  &$0.5$ &$-0.1698$     &$11.01$     &$0.2436$   &$-0.2868$ \\
\hline\hline
\end{tabular}
\caption{The solution of the Einstein constraints of ten PBHLs with spacing of the finest one of $5$ refinement levels equal to $0.0625$. }
\label{tab:PBHLsolution}
\end{table*}

\begin{figure}[]
\begin{center}
\includegraphics[scale=0.35]{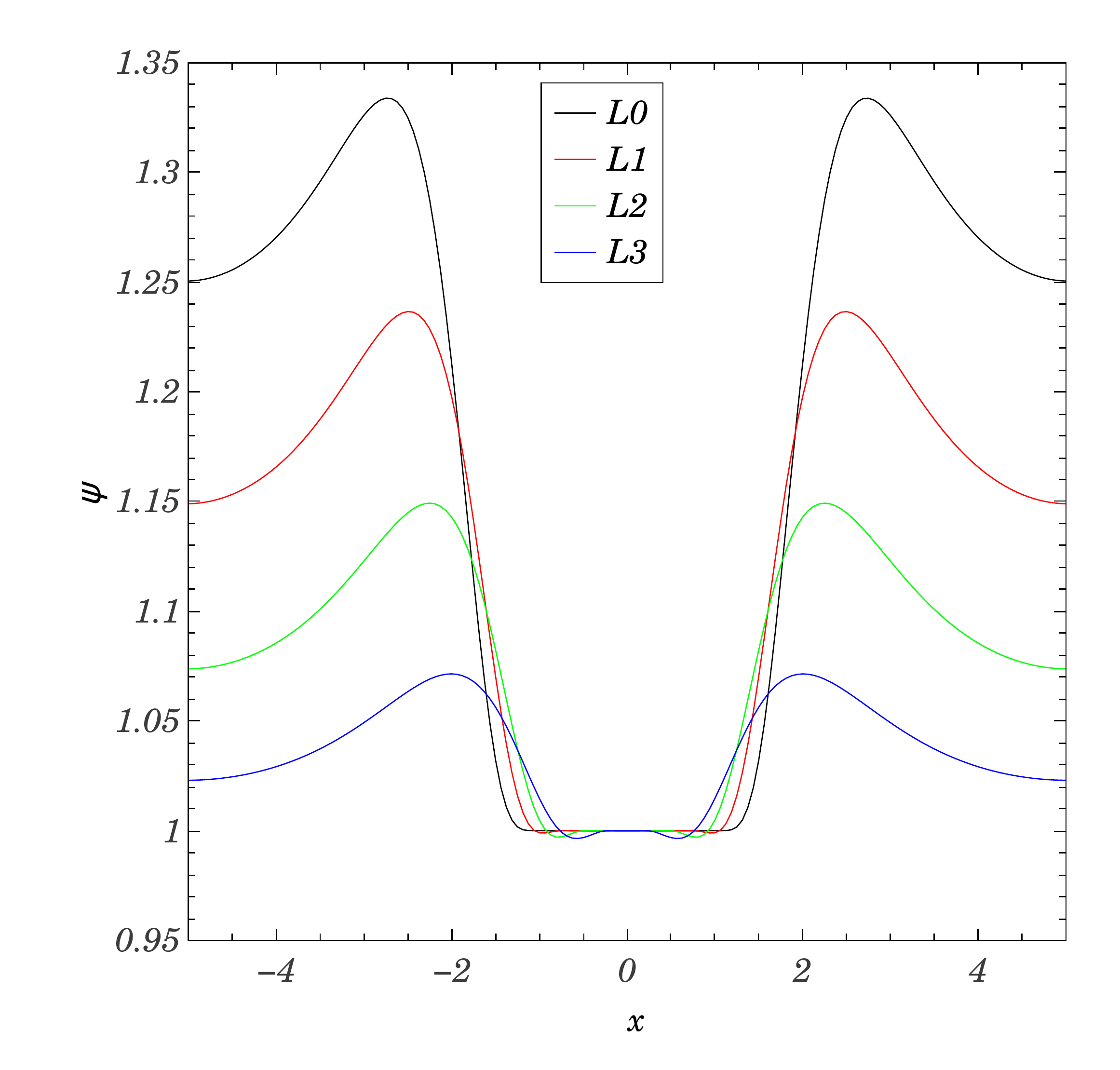}
\includegraphics[scale=0.35]{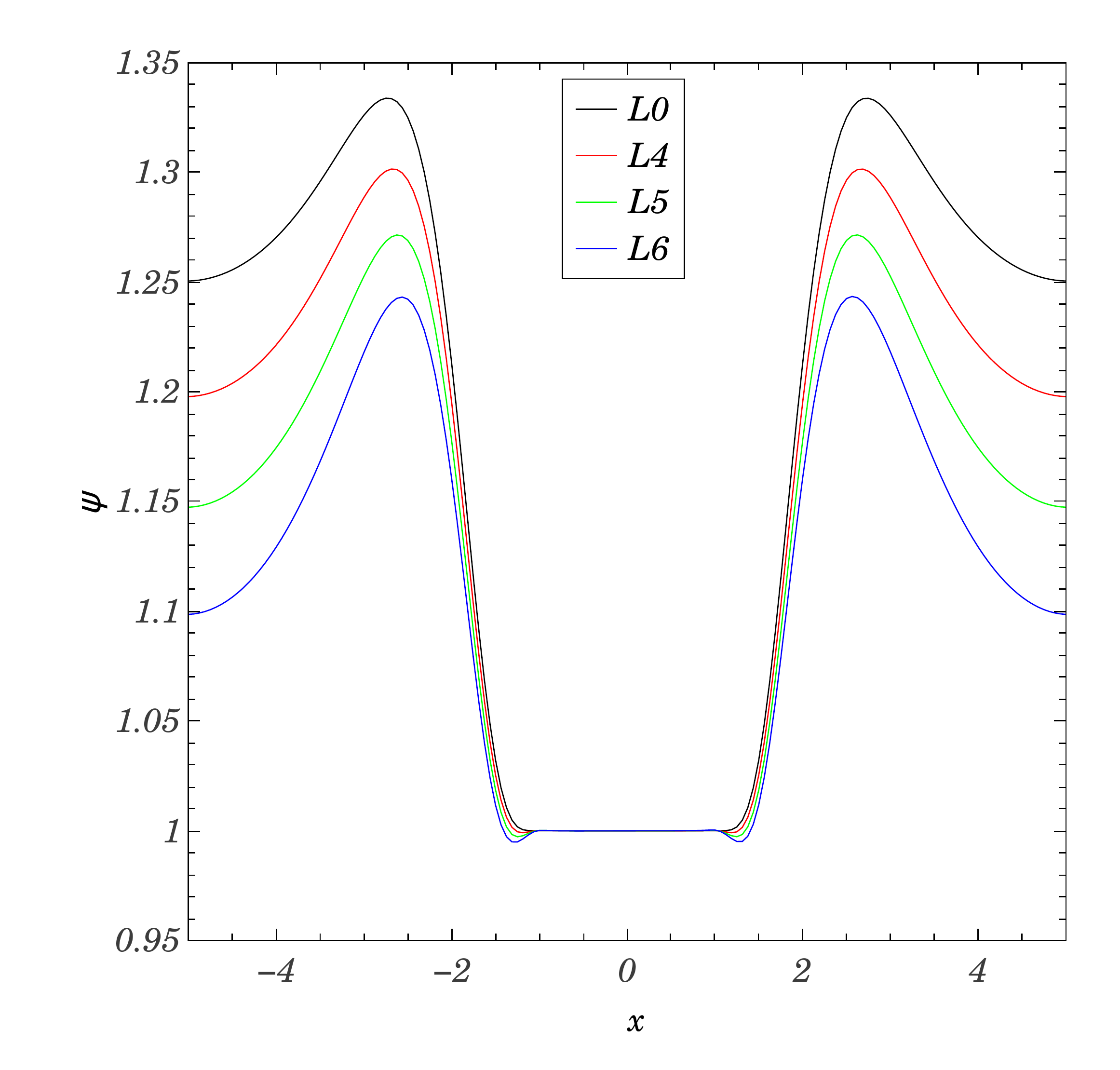}
\includegraphics[scale=0.35]{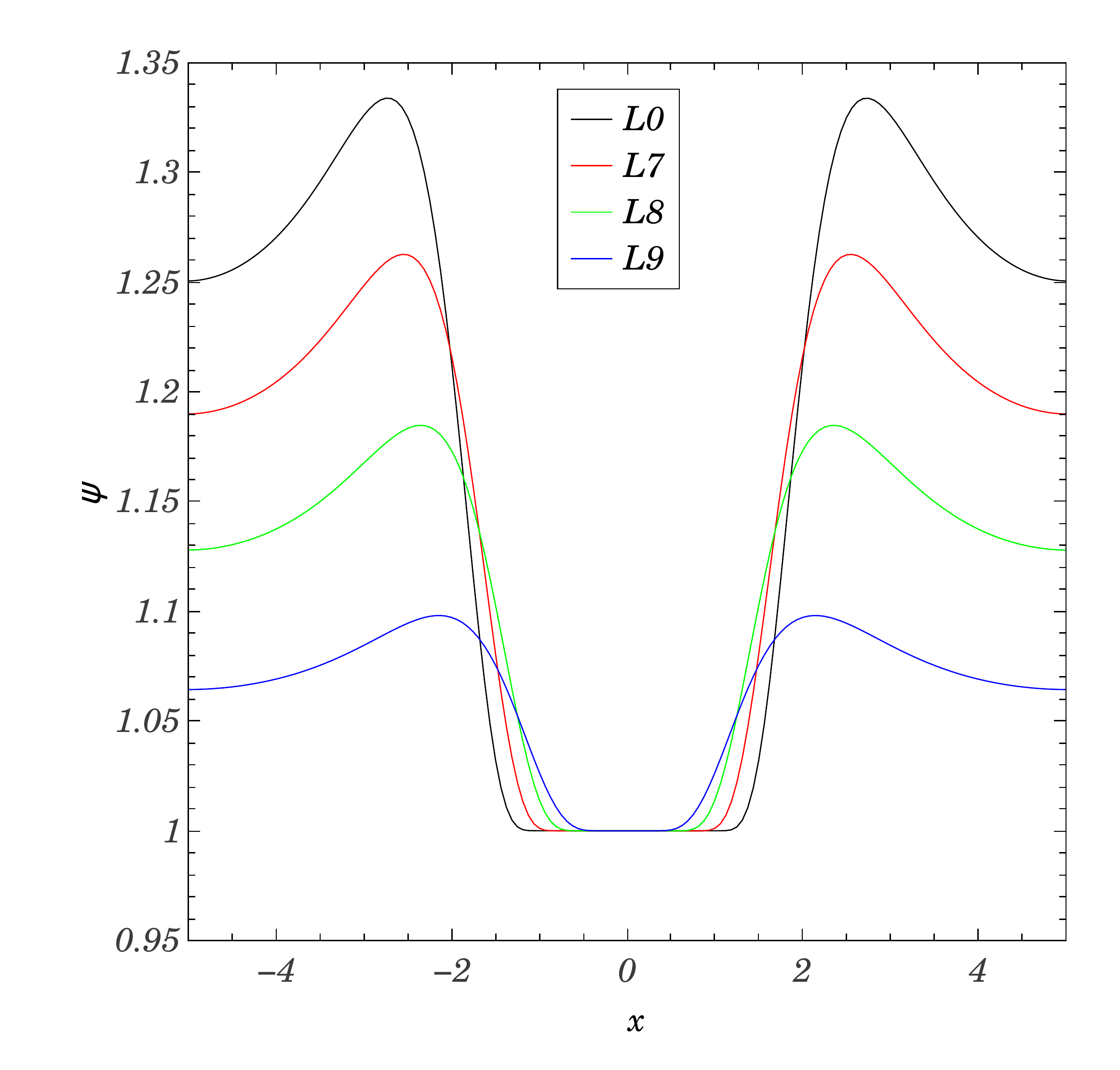}
\includegraphics[scale=0.35]{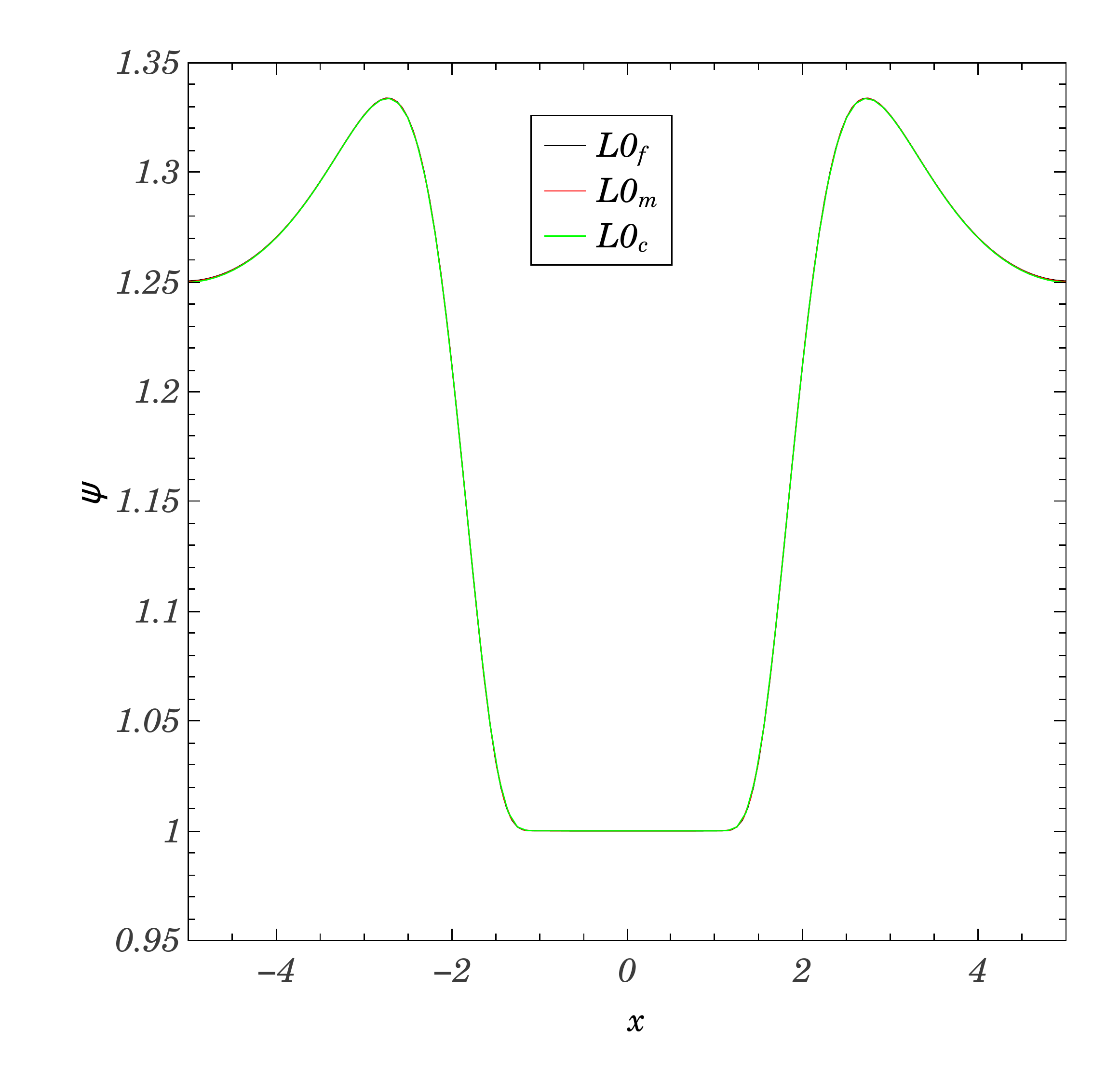}
\end{center}
\caption{$\psi$ of PBHLs as a function of position.}
\label{fig:psi}
\end{figure}

\begin{figure}[]
\begin{center}
\includegraphics[scale=0.35]{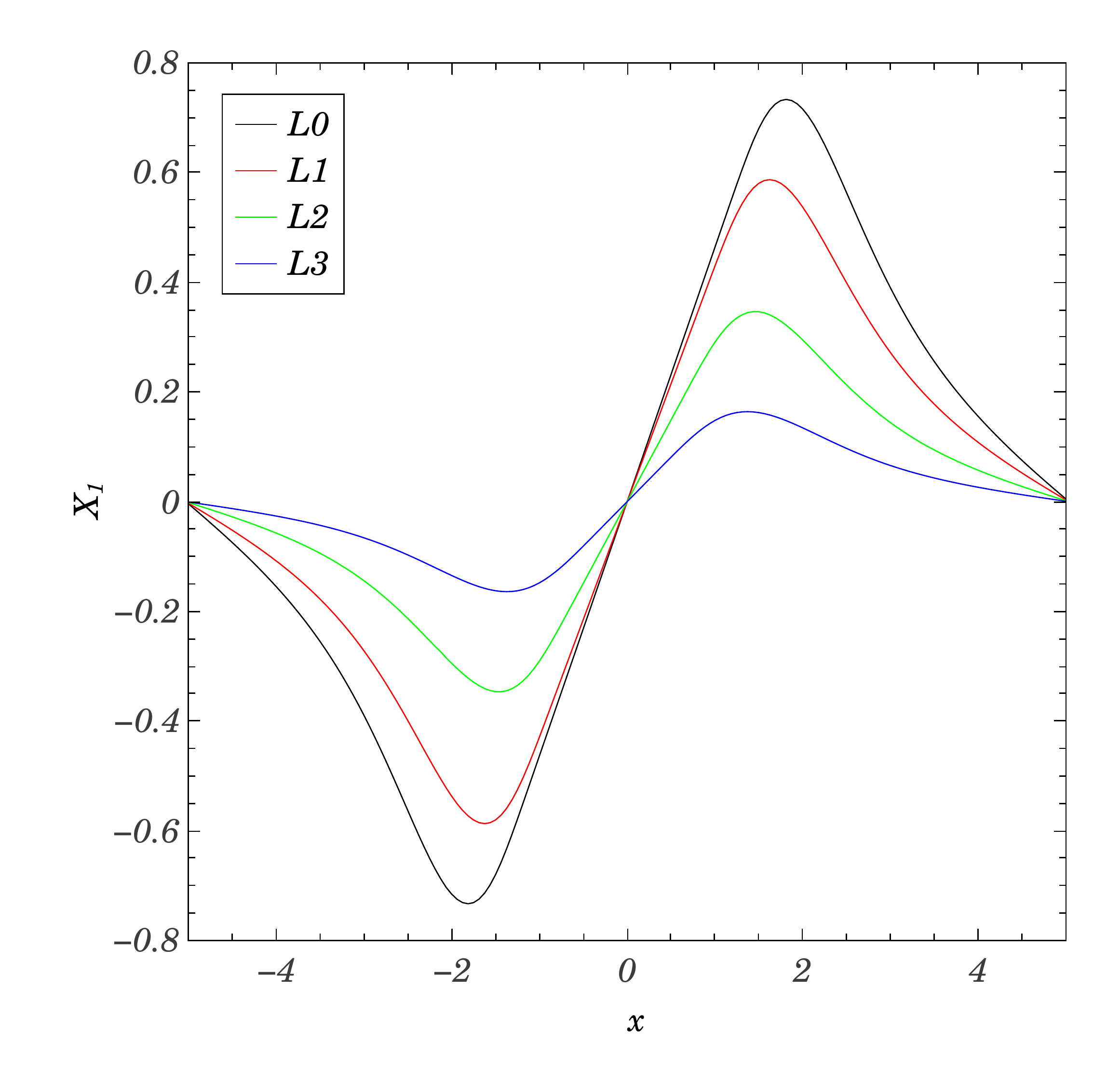}
\includegraphics[scale=0.35]{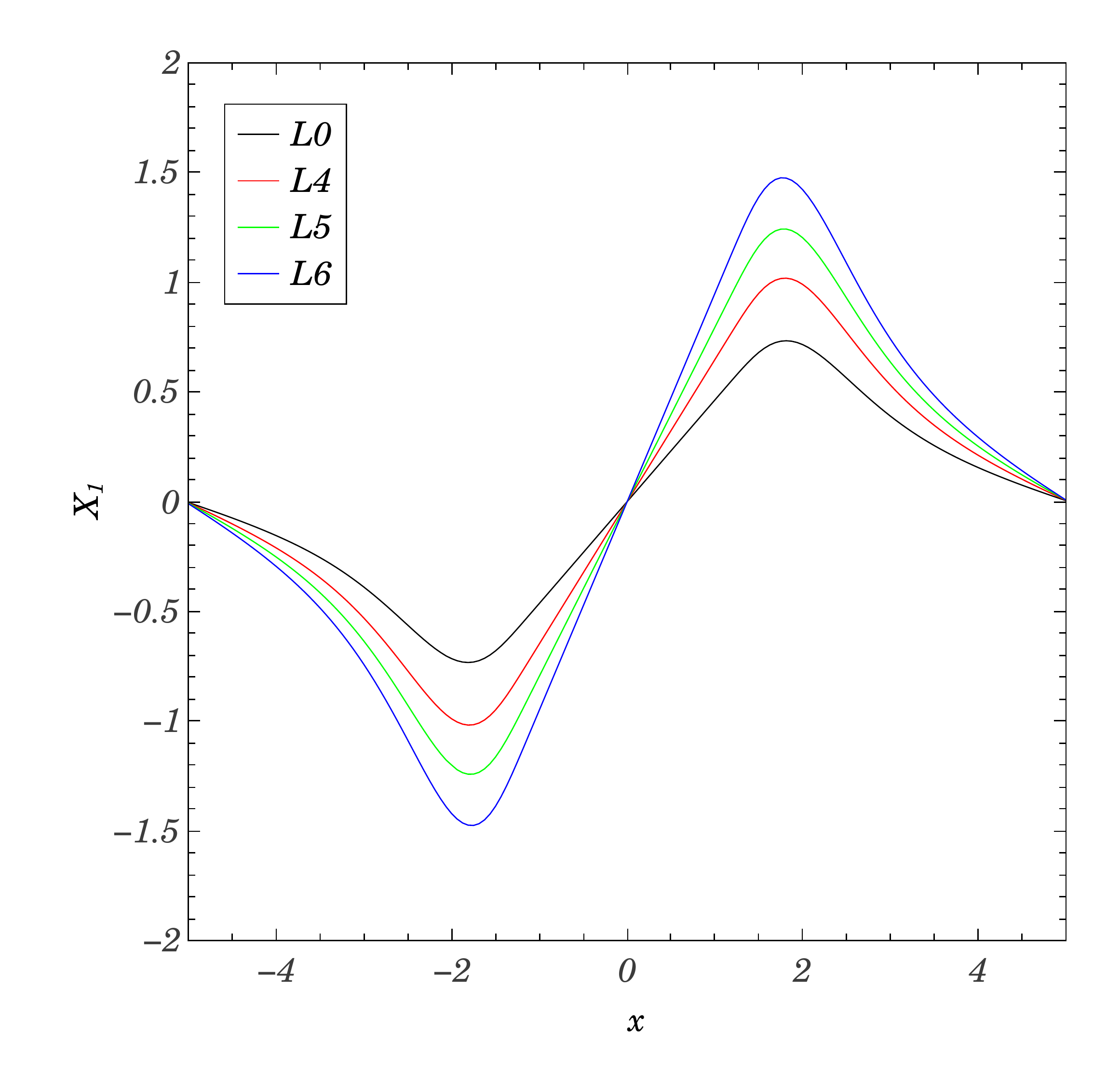}
\includegraphics[scale=0.35]{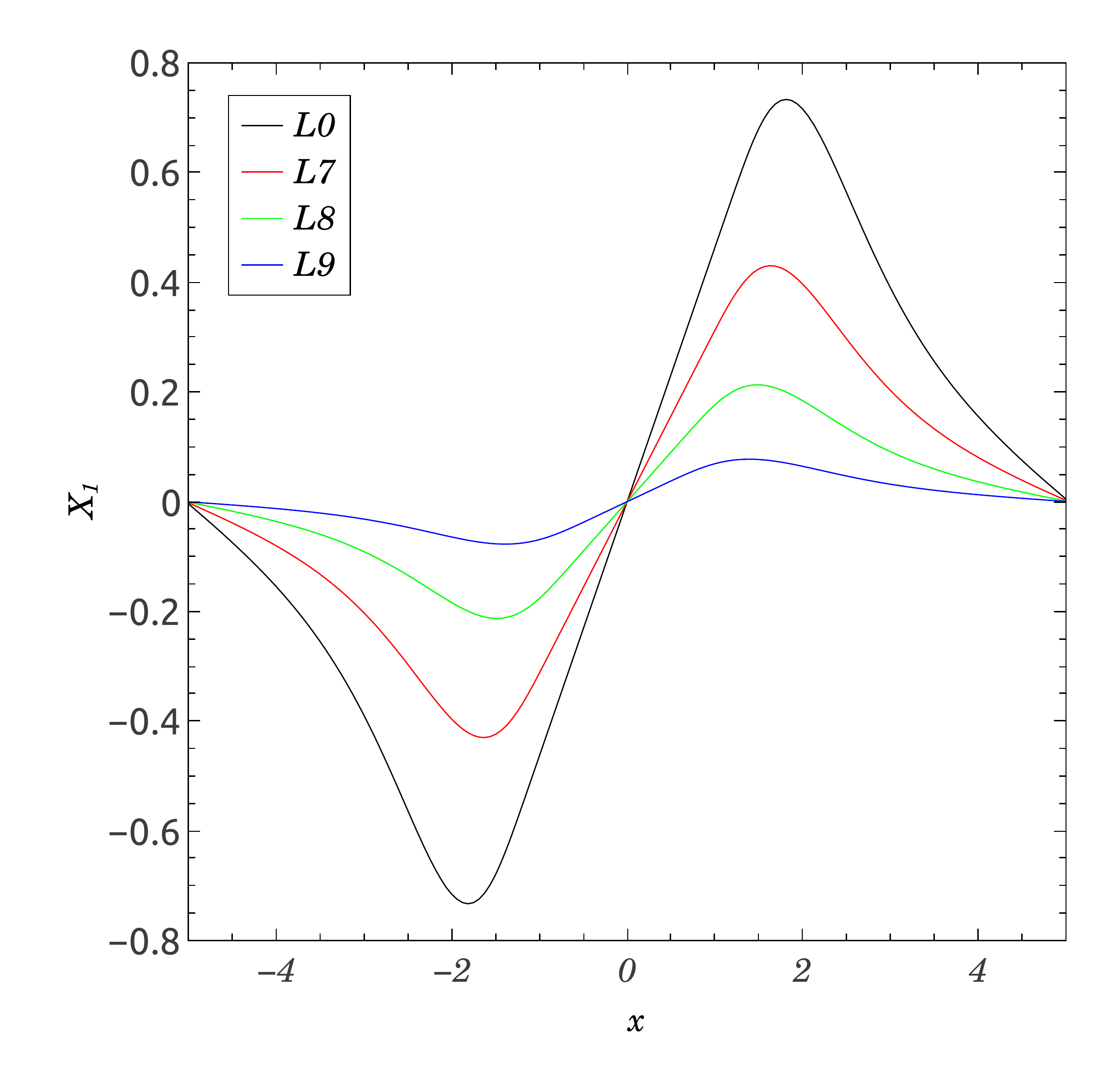}
\includegraphics[scale=0.35]{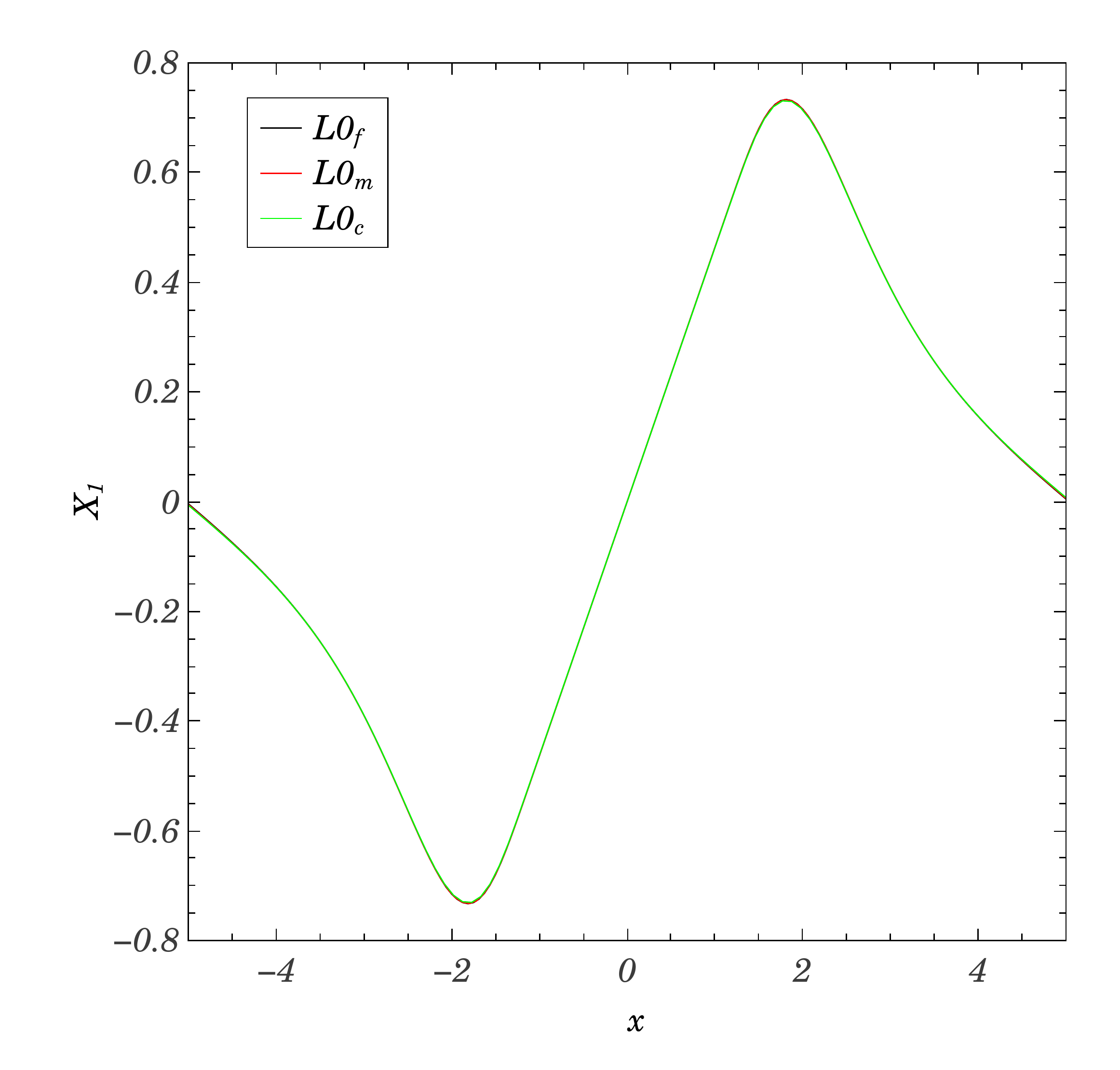}
\end{center}
\caption{$X_1$ of PBHLs as a function of position.}
\label{fig:x}
\end{figure}

\begin{figure}[]
\begin{center}
\includegraphics[scale=0.3]{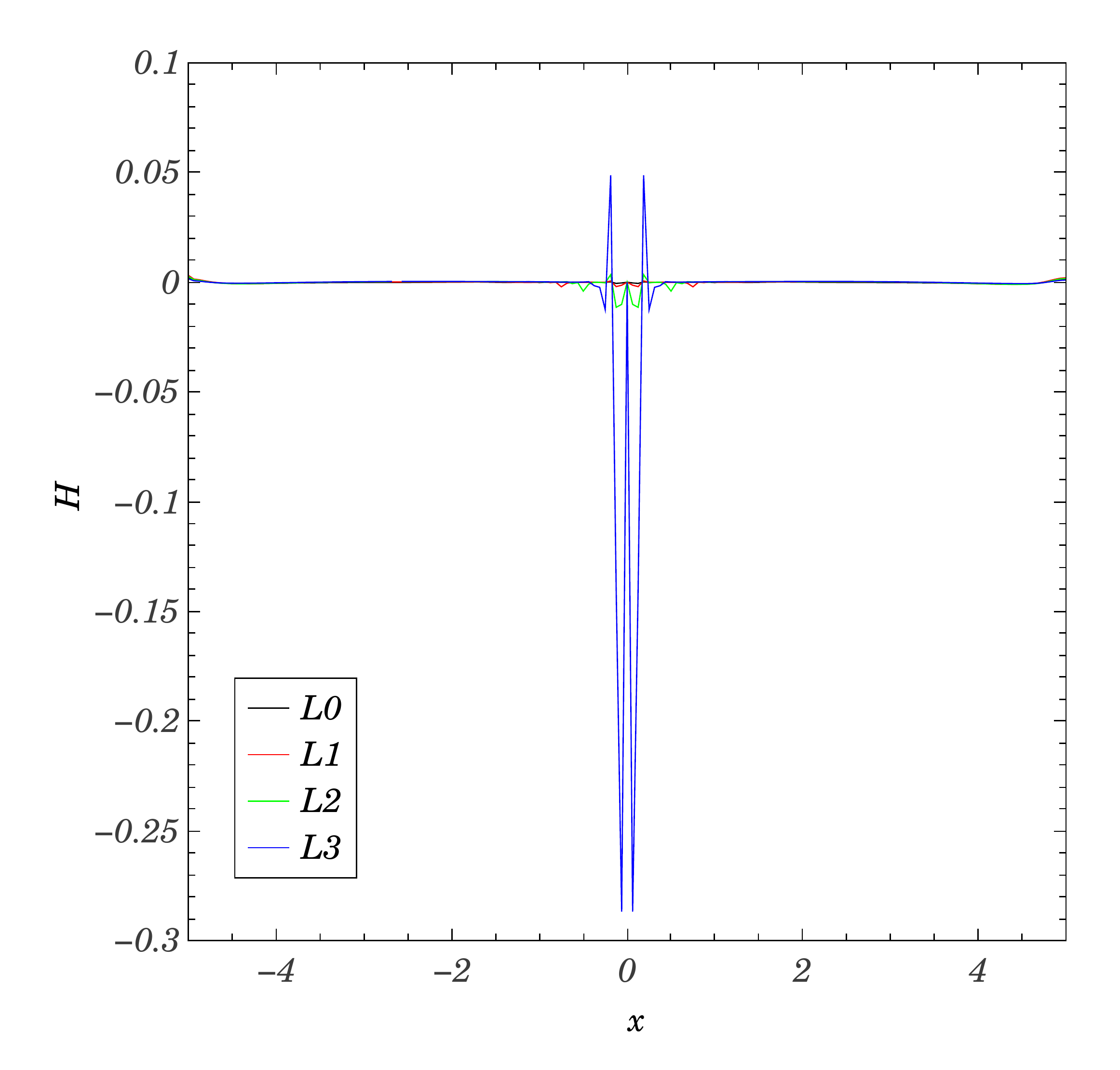}
\includegraphics[scale=0.3]{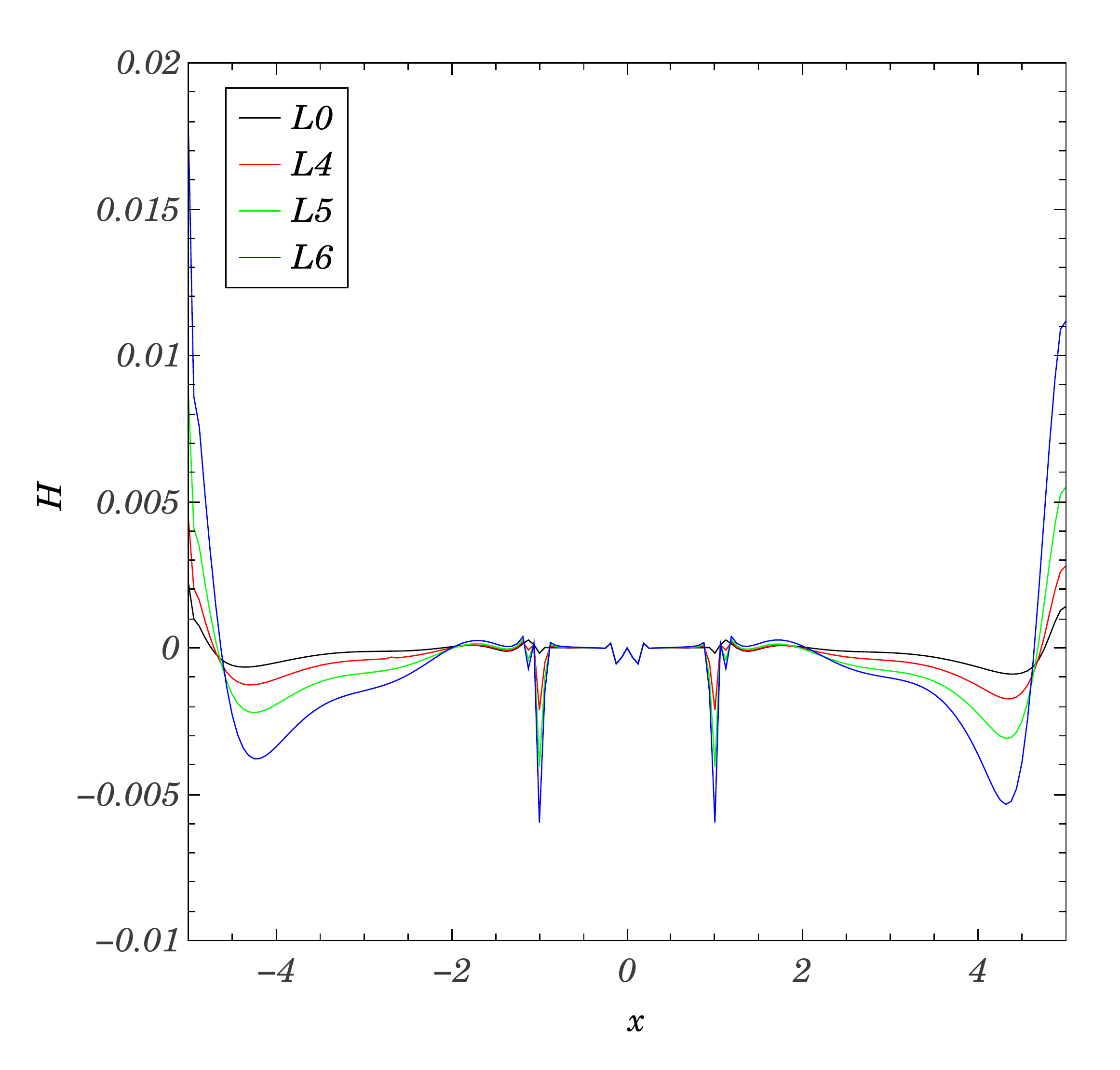}
\includegraphics[scale=0.3]{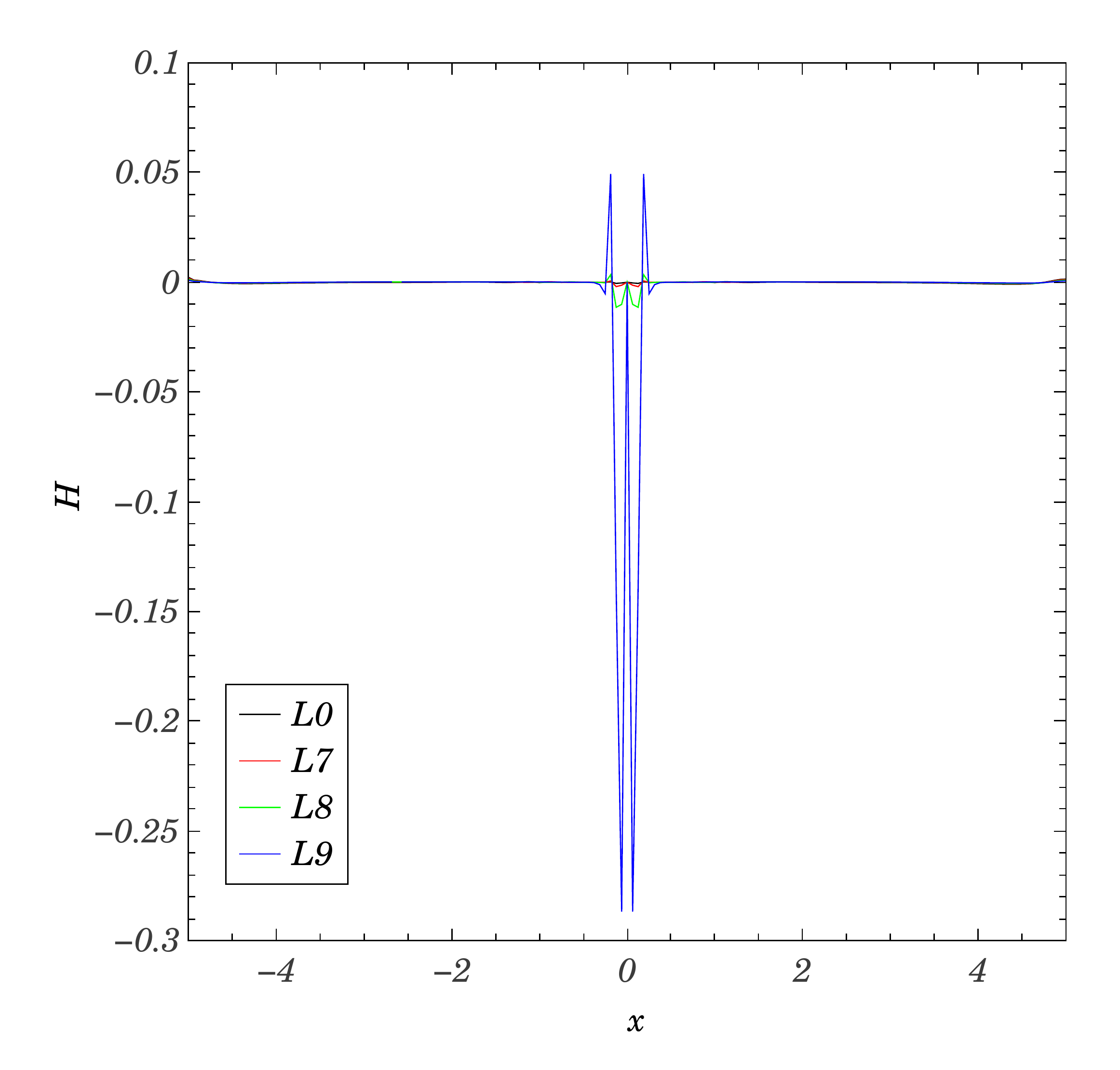}
\includegraphics[scale=0.3]{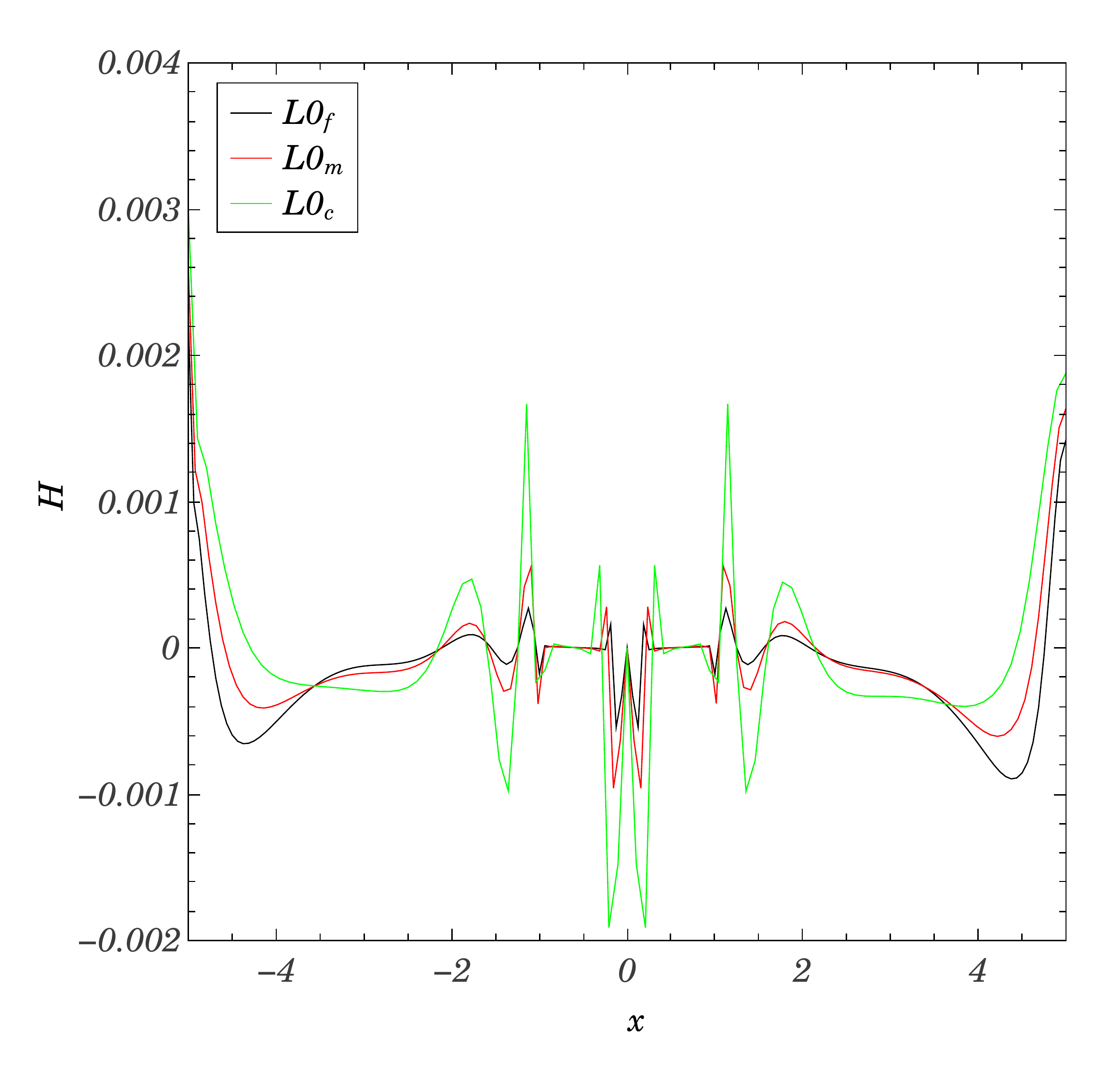}
\end{center}
\caption{The initial Hamiltonian constraint violation of PBHLs.}
\label{fig:H}
\end{figure}

In Fig.~\ref{fig:psi}, the $\Psi=\psi>1$ at the boundary means that each PBH is gravitationally coupled to its neighbors; the boundary $\psi$ of L0 is larger than that of L9 just because the PBH with larger mass will be gravitationally coupled to its neighbors more tightly; the boundary $\psi$ of L0 is larger than that of L6, which means the matter surrounding the PBH suppresses the gravitational correlation among PBHs; the boundary $\psi$ of L0 is larger than that of L3, which is consistent with above two cases.
In Fig.~\ref{fig:x}, we give the solutions of $X_1$ of PBHLs.
In Fig.~\ref{fig:H}, we shows the initial Hamiltonian constraint violation of PBHLs.

The last plot in Fig.~\ref{fig:psi}, Fig.~\ref{fig:x} and Fig.~\ref{fig:H} is a convergence test for solving the initial data and shows the $\psi$, $X_1$ and $H$ of L0 with spacing of the finest one of $5$ refinement levels equal to $0.0625$ (L$0_f$), $0.078125$ (L$0_m$) and $0.1041667$ (L$0_c$) respectively.

\section{Time Evolution of Primordial Black Hole Lattices}
\label{evolution}

Given the initial data of PBHLs, we can simulate the time evolution of PBHLs. Due to the negative trace of the extrinsic curvature $K_{ij}$ around the boundary, we can expect an expansion of PBHLs during the time evolution. Meanwhile, since the motion of PBHs caused by the expansion of PBHLs occurs at speeds close to that of light, we also expect the emission of GWs during the time evolution.
\subsection{Expansion of Primordial Black Hole Lattices}
\label{expansion}

In order to circumvent the central puncture, here we will study the expansion of PBHLs through the rescaling of the proper length of lattice cells' edge with proper time. The proper time at anywhere is given by integrating the corresponding lapse function $\alpha(t,x)$ from the beginning of simulation
\begin{equation}
 \tau(t,x)=\int_0^t\alpha(t',x)dt',
\end{equation}
where the evolution of $\alpha$ is given by
\begin{equation}
\partial_t\alpha-\beta^i\partial_i\alpha=-2K\alpha.
\end{equation}
And the proper length measured in a \textsf{cctk\_delta\_time} (or a timestep) is then given by
\begin{equation}
 D(\tau)=\int_{\gamma_\tau}[(-\alpha^2(\tau,\ell)+\beta^2(\tau,\ell))(\partial_\ell t)^2+2\beta_i(\tau,\ell)\partial_\ell t\partial_\ell x^i+\gamma_{ij}(\tau,\ell)\partial_\ell x^i\partial_\ell x^j]^{1/2}d\ell,
\end{equation}
where $\beta_i(\tau,x)$ is the shift vector obeying a Gamma driver and $\gamma_\tau$ is a constant-$\tau$ edge parameterized by $\ell$.

\begin{figure}[]
\begin{center}
\includegraphics[scale=0.35]{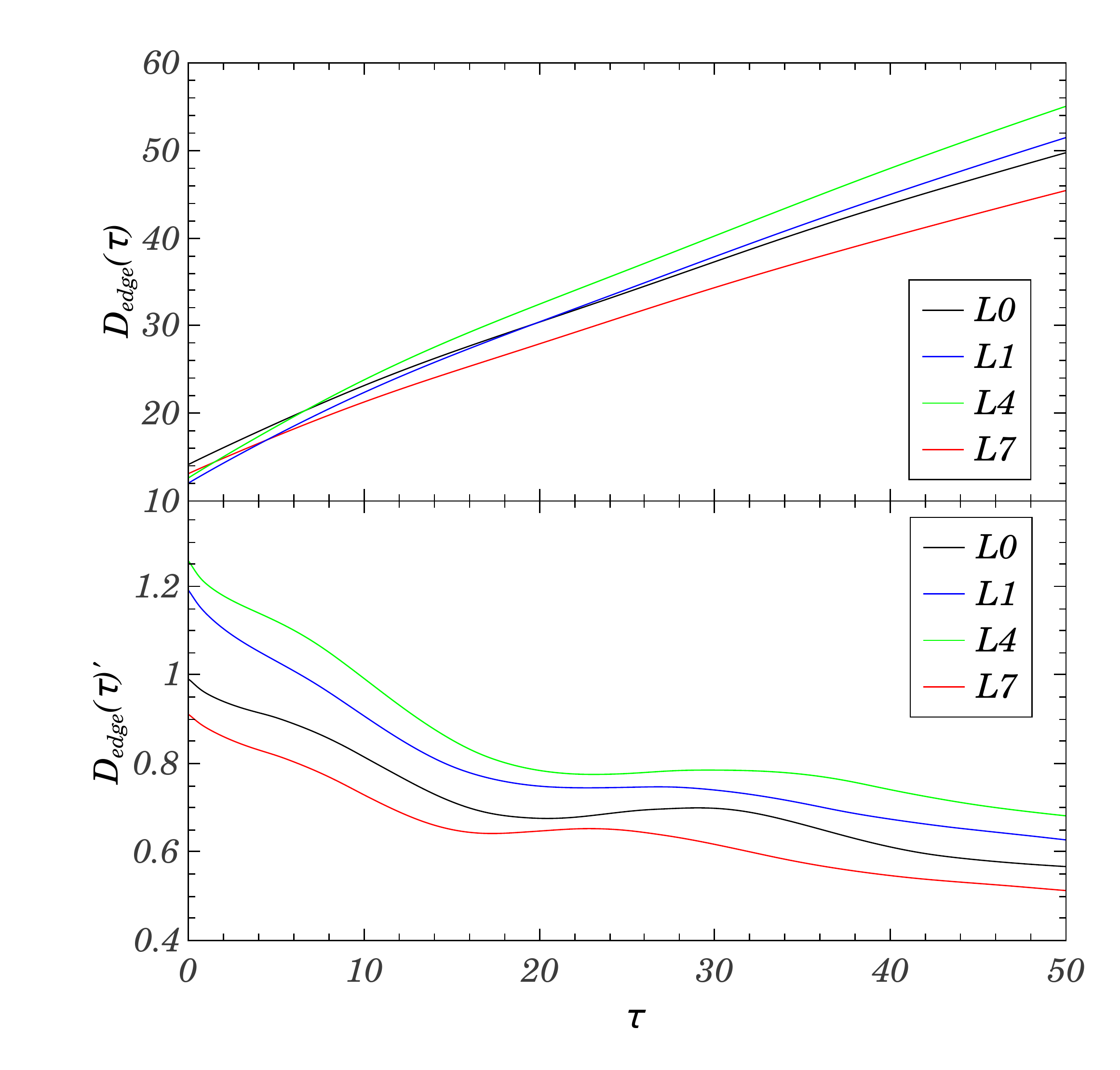}
\includegraphics[scale=0.35]{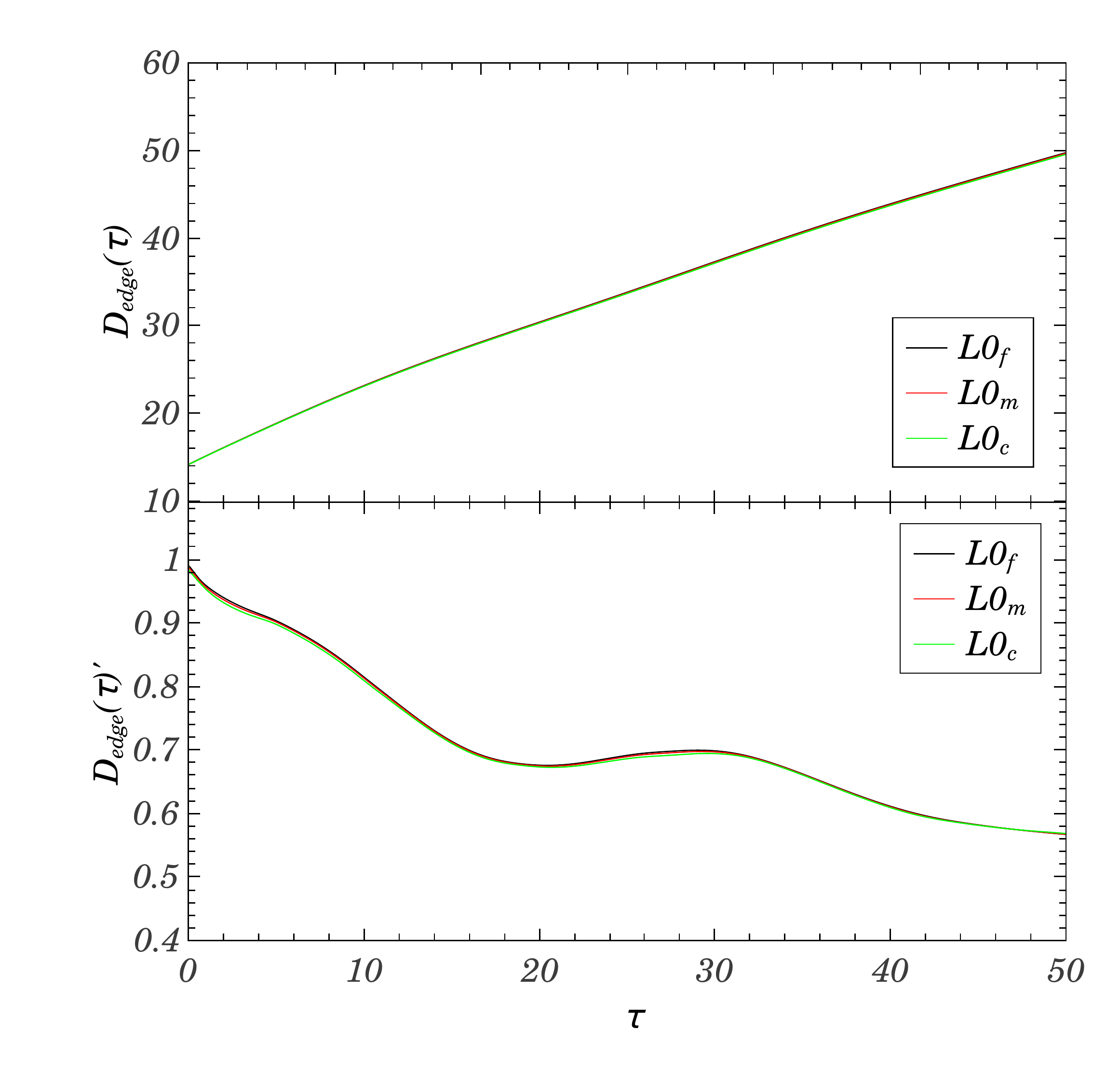}
\end{center}
\caption{Expansion of four PBHLs with the smallest initial Hamiltonian constraint violation in each group.}
\label{fig:expansion}
\end{figure}

Fig.~\ref{fig:expansion} shows the expansion of four PBHLs with the smallest initial Hamiltonian constraint violation in each group, where a prime represents a derivative with respect to the proper time $\tau$. We can see that $L0$ and $L1$ evolve differently even though they share a same $\rho_t$. That is to say $f_{\mathrm{PBH}}$ will play an important role during the evolution of PBHLs. Comparing $L4$ to $L0$ and $L1$, we find that the PBHL with a larger $\rho_t$ no matter due to an extra $\rho_m$ or a larger $m_{\mathrm{PBH}}$ expands faster. There is no intersection point between the evolutions of $D_{\mathrm{edge}}(\tau)'$, which means a PBHL with a larger $-K_c$ will keep expanding faster for ever. Finally, we also find that the motion of PBHs caused by the expansion of PBHLs does occur at speeds close to that of light.

\begin{figure}[]
\begin{center}
\includegraphics[scale=0.35]{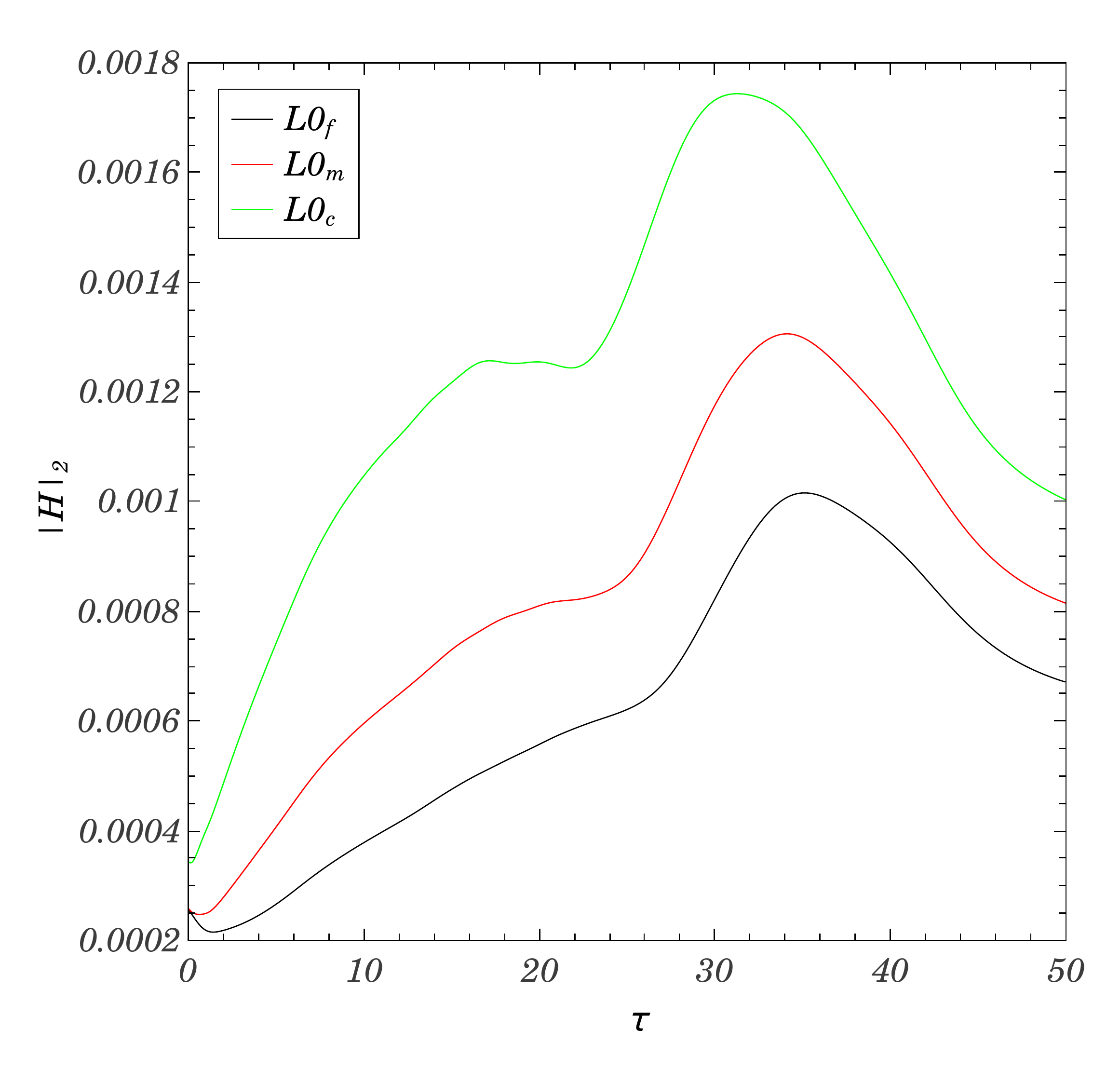}
\includegraphics[scale=0.35]{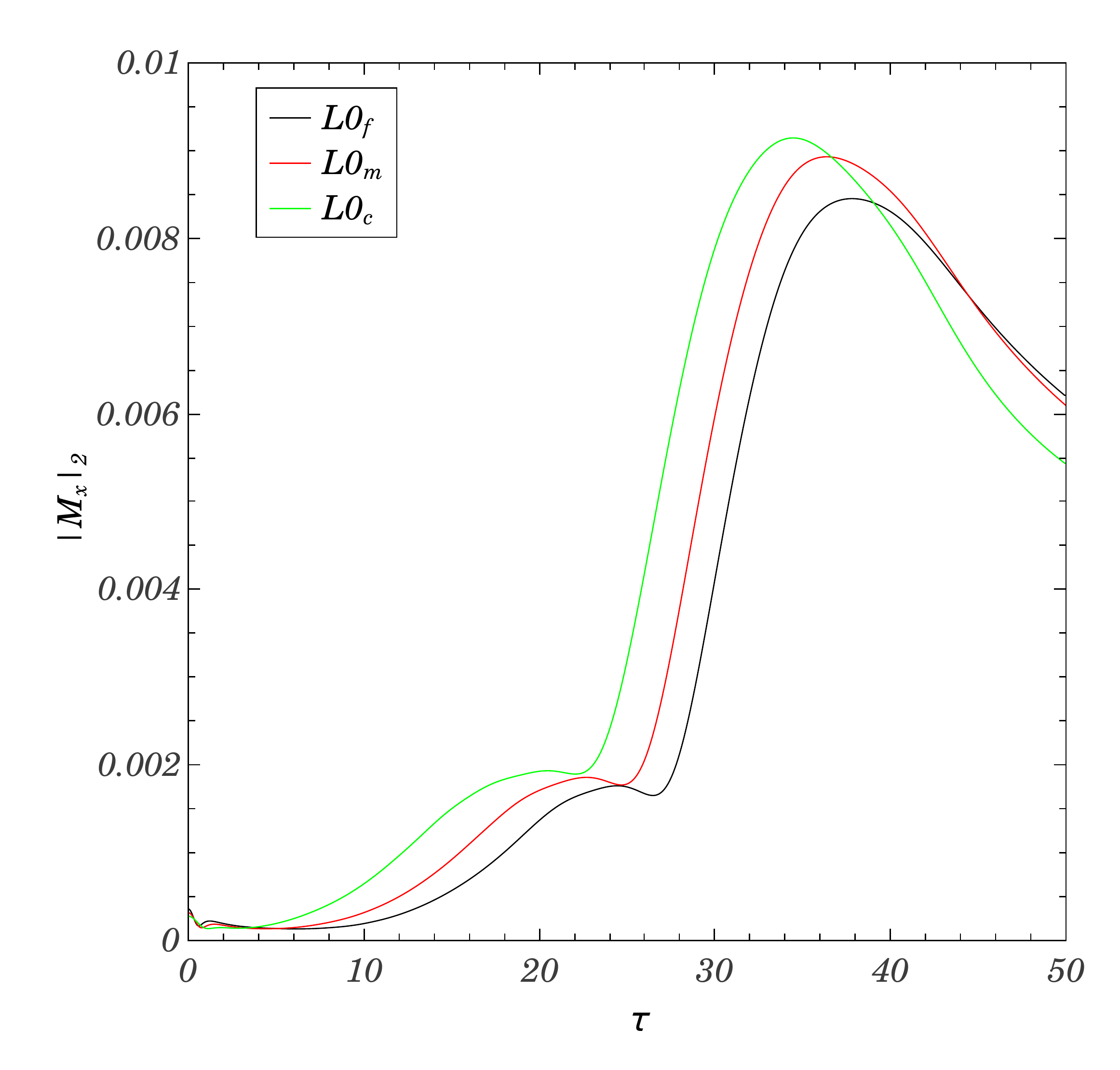}
\end{center}
\caption{$L_2$ norms of the Hamiltonian and momentum constraint for $L0$ with spacing of the finest one of $5$ refinement levels equal to $0.0625$, $0.078125$ and $0.1041667$ respectively.}
\label{fig:constraint}
\end{figure}

The right plot in Fig.~\ref{fig:expansion} shows the expansion of $L0$ and  Fig.~\ref{fig:constraint} shows the $L_2$ norms of the Hamiltonian and momentum constraint for $L0$ with spacing of the finest one of $5$ refinement levels equal to $0.0625$, $0.078125$ and $0.1041667$ respectively. They serve as a convergence test for evolving PBHLs.

\subsection{Gravitational Waves in Primordial Black Hole Lattices}
\label{production}
We have shown that the motion of PBHs caused by the expansion of PBHLs occurs at speeds close to that of light. Here we will use both analytical estimates and numerical simulations to cross check the production of GWs in expanding PBHLs.

\subsubsection{Theoretical estimation}
The total energy radiated by one PBH in an expanding PBHL between $\tau$ and $\tau+\Delta\tau$ can be considered as the gravitational radiation from an accelerated mass estimated by \cite{Maggiore:1900zz}
\begin{eqnarray}
\label{theoretical}
\nonumber
E&=&\frac{1}{2\pi^2}\int d\Omega~\Lambda_{ij,kl}(\hat{\textbf{n}})\int_0^{\infty}d\omega~ \omega^2\tilde{T}_{ij}(\omega,\omega\hat{\textbf{n}})\tilde{T}_{kl}^*(\omega,\omega\hat{\textbf{n}})\\
&\approx & \frac{\gamma^2(\tau)m_{\mathrm{PBH}}^2}{2\pi}~T~ [v(\tau)+v(\tau+\Delta\tau)]^2~[v(\tau)- v(\tau+\Delta\tau)]^2  \int d\Omega\frac{\sin^4\theta}{[1- v(\tau)\cos\theta]^2}
\end{eqnarray}
where the energy-momentum tensor is
\begin{equation}
\tilde{T}^{ij}(\omega,\omega\hat{\textbf{n}})=2\pi\delta(\omega-\omega_0)\frac{-i~ m_{\mathrm{PBH}}}{\omega}\left[\frac{\gamma(\tau) v^i(\tau)v^j(\tau)}{1- v(\tau)\cos\theta}-\frac{\gamma(\tau+\Delta\tau) v^i(\tau+\Delta\tau)v^j(\tau+\Delta\tau)}{1- v(\tau+\Delta\tau)\cos\theta}\right],
\end{equation}
the Lambda tensor $\Lambda_{ij,kl}(\hat{\textbf{n}})$ is
\begin{equation}
\Lambda_{ij,kl}(\hat{\textbf{n}})=\delta_{ik}\delta_{jk}-\frac{1}{2}\delta_{ij}\delta_{kl}-n_jn_l\delta_{ik}-n_in_k\delta_{jl}+\frac{1}{2}n_kn_l\delta_{ij}+\frac{1}{2}n_in_j\delta_{kl}+\frac{1}{2}n_in_jn_kn_l,
\end{equation}
the scalar product of the direction of gravitational radiation $\hat{\textbf{n}}$ and the velocity of PBH $\textbf{v}$ is $n_iv^i=v\cos\theta$,
$\omega_0$ is the frequency of the gravitational radiation at $\tau$,
$T=2\pi \delta(0)\approx\Delta\tau$ and $\gamma=(1-v^2)^{-1/2}$.
The term of $[v(\tau)-v(\tau+\Delta\tau)]^2$ predicts that, in our PBHLs, the total energy radiated when $\tau\lesssim14$ is larger than that when $\tau\gtrsim14$ because $|D_{\mathrm{edge}}(\tau\lesssim14)''|$ is much larger than $|D_{\mathrm{edge}}(\tau\gtrsim14)''|$ as show in Fig.~\ref{fig:expansion}.
The term of $[v(\tau)+v(\tau+\Delta\tau)]^2$ predicts that when $\tau\lesssim14$ the total energy radiated in L4 should be larger than that in L0 and L7 because $D_{\mathrm{edge}}(\tau)'$ of L4 is larger than that of L0 and L7. Similarly, according to the initial distribution of $K=K_cW(r)$, the total energy radiated in PBHLs should not be uniform and the locations far away from the center are full of more radiation energy.

\subsubsection{Simulation results}
\begin{figure}[]
\begin{center}
\includegraphics[scale=0.27]{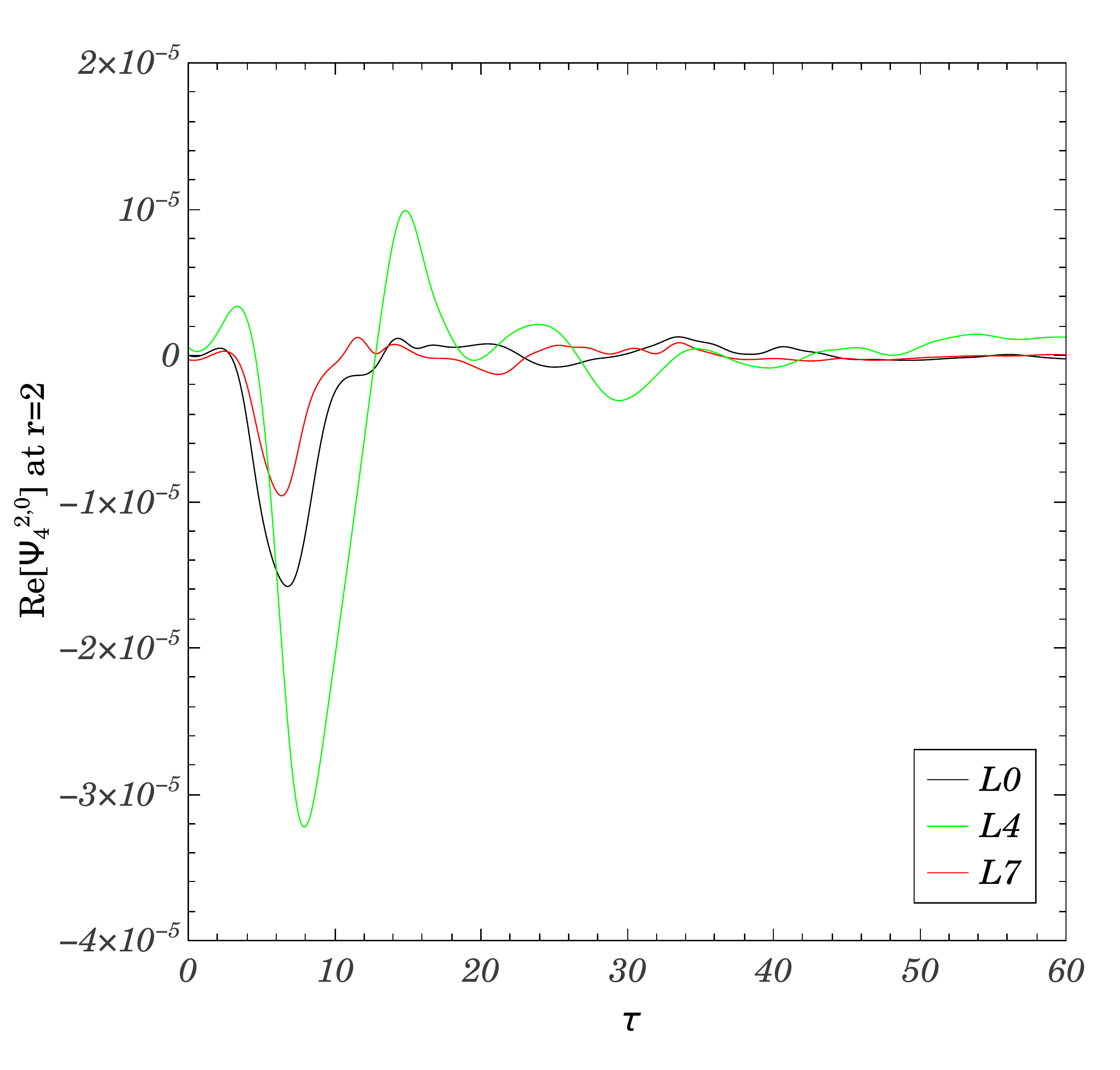}
\includegraphics[scale=0.27]{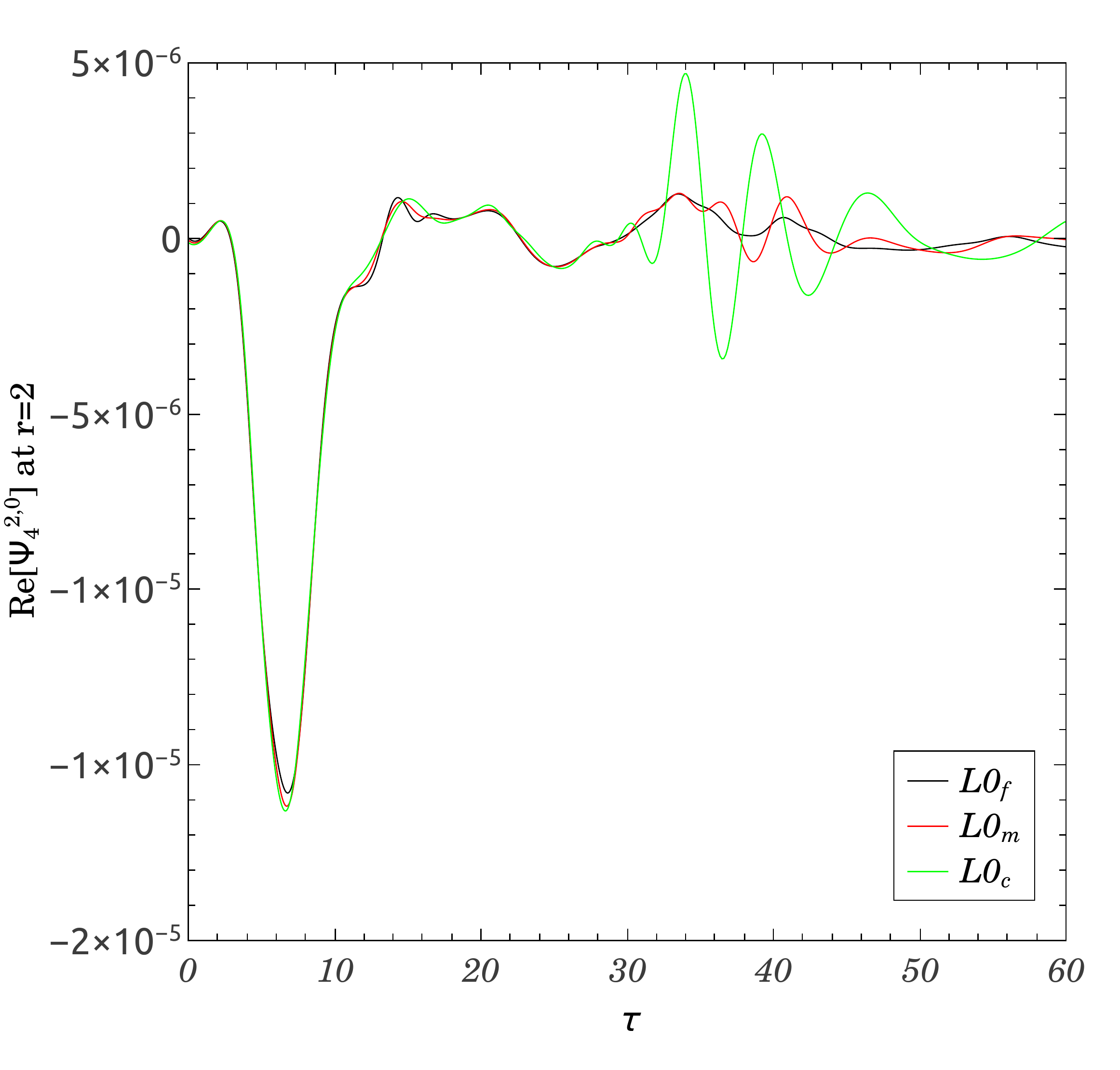}
\includegraphics[scale=0.27]{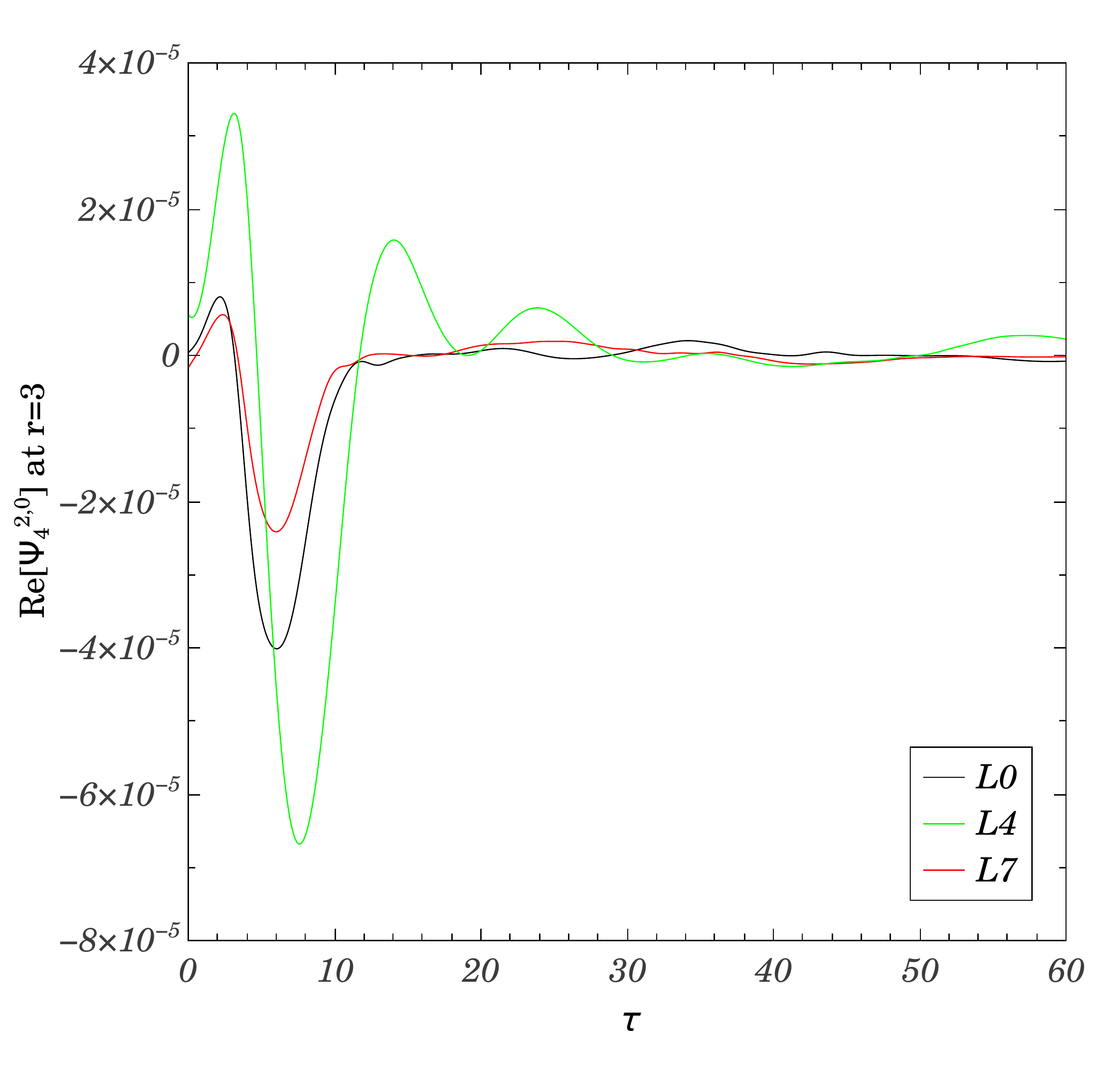}
\includegraphics[scale=0.27]{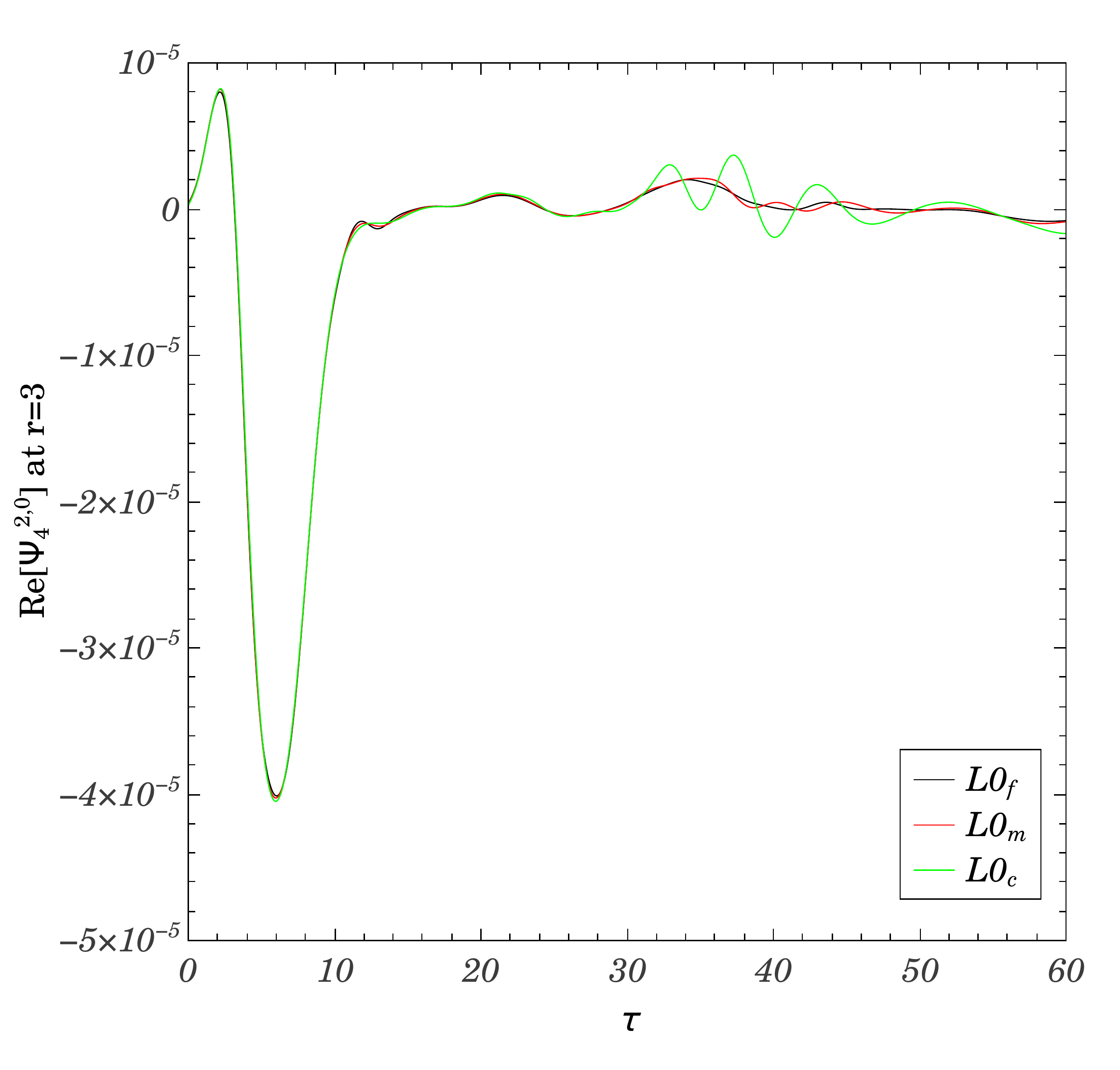}
\includegraphics[scale=0.27]{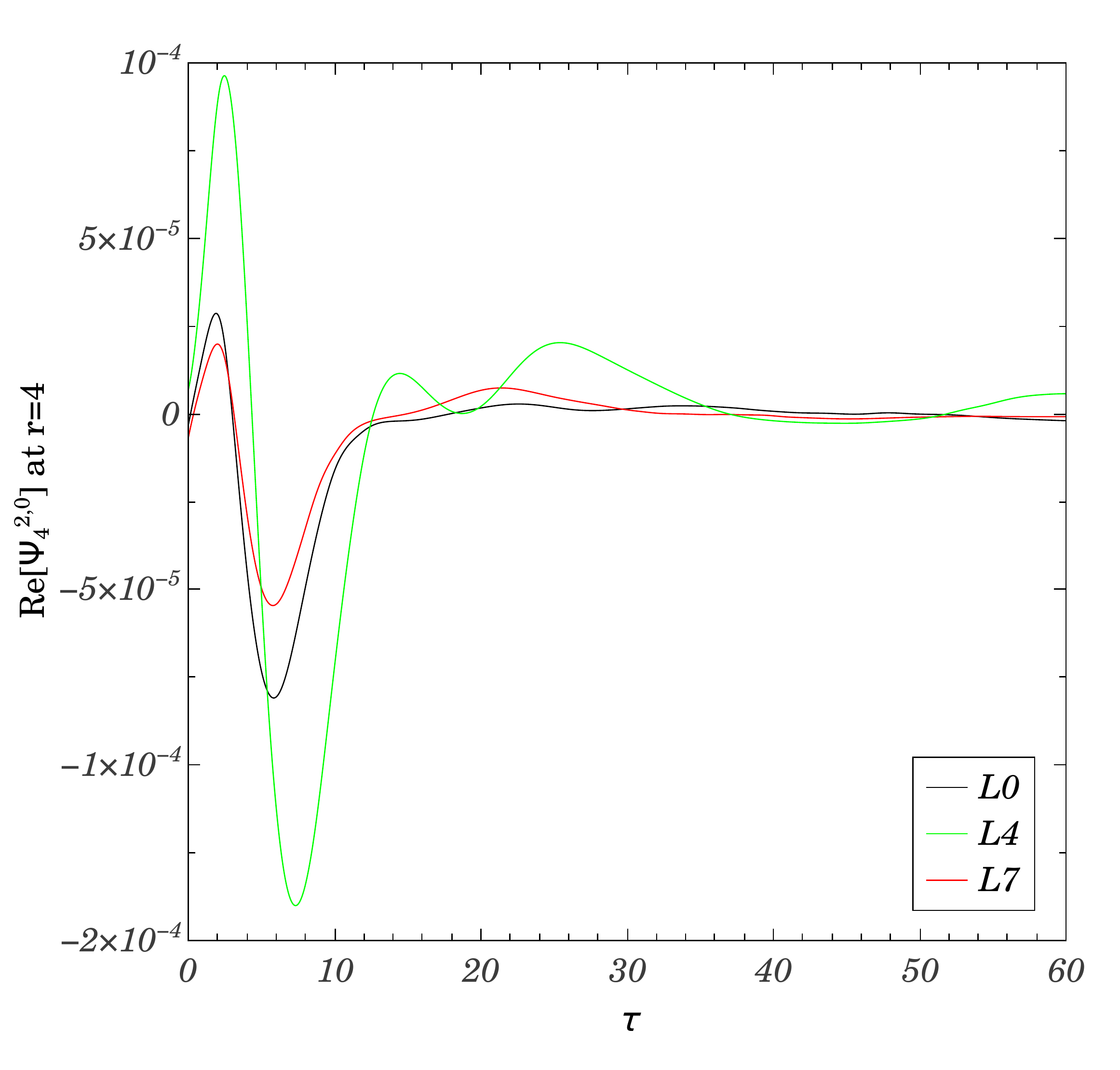}
\includegraphics[scale=0.27]{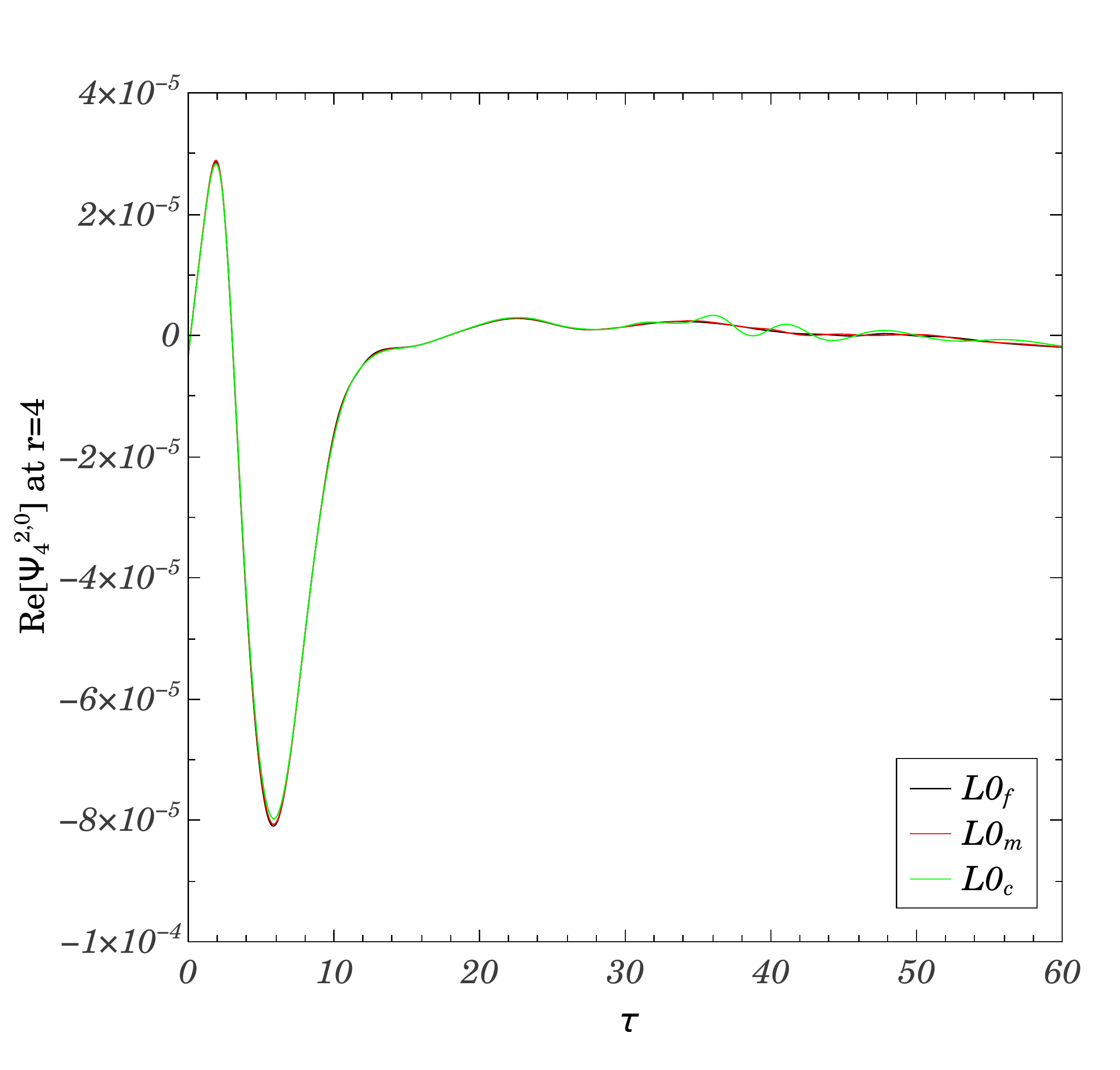}
\includegraphics[scale=0.27]{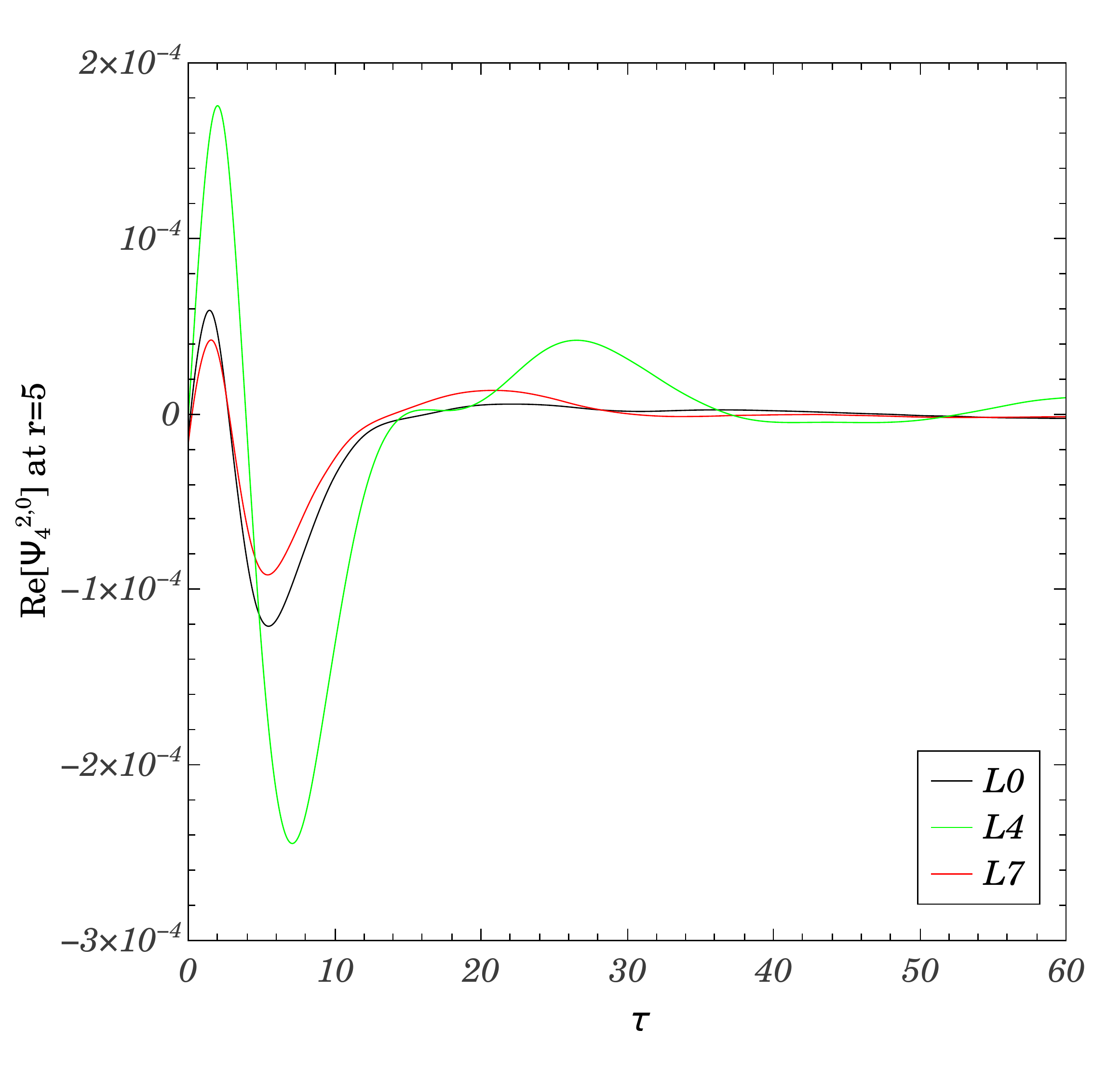}
\includegraphics[scale=0.27]{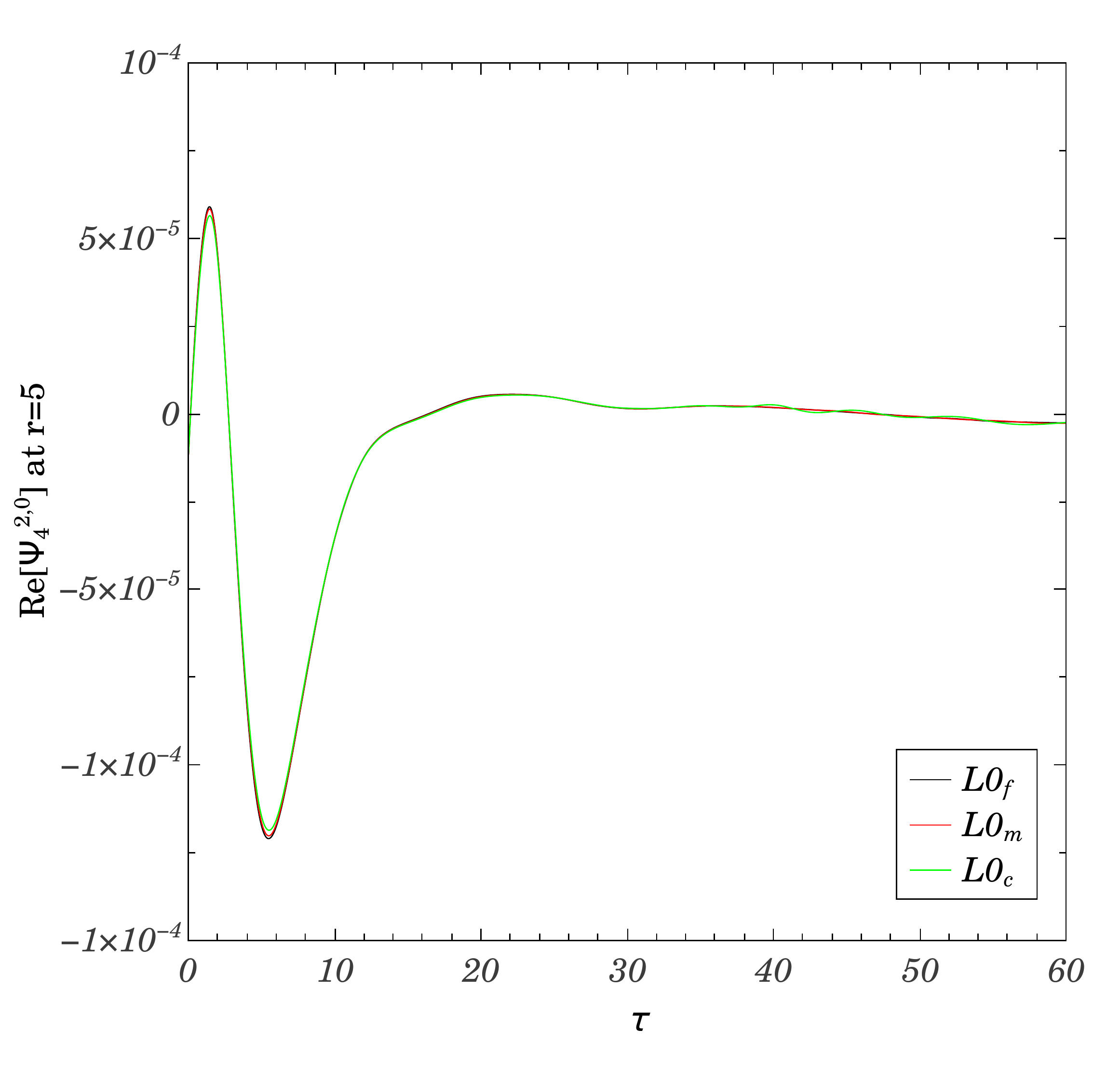}
\end{center}
\caption{A sample of GWs emitted in expanding PBHLs, as estimated by the Newman-Penrose scalar $\Psi_4$.}
\label{fig:GW}
\end{figure}

Fig.~\ref{fig:GW} shows the waveforms of GWs, as estimated by the Newman-Penrose scalar $\Psi_4(\tau)=h''_+-ih''_{\times}$, produced at several distances $r$ in expanding PBHLs with different $\rho_t$. For $\tau\lesssim14$, there is an obvious production of GWs and we also find an obvious deceleration in Fig.~\ref{fig:expansion}; for $\tau\gtrsim14$, the amplitude of waveforms decreases and there is a plateau for $D_{\mathrm{edge}}(\tau)'$ in Fig.~\ref{fig:expansion}. That is to say, there are some gravitational potential energy among PBHs converted to the GWs radiation and Re$[\Psi_4^{2,0}]\propto |D_{\mathrm{edge}}(\tau)''|$. The amplitude of waveforms increases with $\rho_t$ increasing, which also means Re$[\Psi_4^{2,0}]\propto D_{\mathrm{edge}}(\tau)'$ since the $D_{\mathrm{edge}}(\tau)''$ is not sensitive to $\rho_t$ as shown in Fig.~\ref{fig:expansion}. The amplitude of GWs increases with $r$ increasing, which means the expansion of PBHLs is not uniform and the initial distribution of $K=K_cW(r)$ keeps a faster expansion at positions far away from PBHs. All of above features are consistent with theoretical predictions.

The right plots in Fig.~\ref{fig:GW} show the waveforms of GWs produced at several distances $r$ in $L0$. The oscillations due to numerical error at $\tau\approx35$ in $L0_c$ disappear in $L0_f$, which guarantees the other oscillations in Fig.~\ref{fig:GW} are the waveforms of GWs produced in expanding PBHLs.

\section{Summary and discussion}
\label{summary}
In this paper, we first use the wide-used \textsf{Einstein Toolkit} to solve the Einstein constraints of PBHLs with different value of $m_{\mathrm{PBH}}$ and $f_{\mathrm{PBH}}$ which is determined by the surrounding flat-distributed dust with energy density $\rho_m$. From the solutions, we find that a smaller PBH mass $m_{\mathrm{PBH}}$ gives a shorter initial proper cubic edge $D_{\mathrm{edge}}(\tau=0)$ and leads a lower initial expansion rate $-K_c$; a higher matter energy density $\rho_m$ has a higher initial expansion rate $-K_c$ but produces a shorter initial proper cubic edge $D_{\mathrm{edge}}(\tau=0)$; a larger PBH mass $m_{\mathrm{PBH}}$ and a smaller matter energy density $\rho_m$ still produce a larger initial proper cubic edge $D_{\mathrm{edge}}(\tau=0)$ but lead a lower initial expansion rate $-K_c$; a smaller PBH mass $m_{\mathrm{PBH}}$ violates the Hamiltonian constraint more severely;
the PBH with larger mass will be gravitationally coupled to its neighbors more tightly; the matter surrounding the PBH suppresses the gravitational correlation among PBHs.

And then we simulate the expansion of PBHLs with the smallest initial Hamiltonian constraint violation in each group. From the evolutions, we find that $f_{\mathrm{PBH}}$ plays an important role during the evolution of PBHLs; the PBHL with a larger $\rho_t$ no matter due to an extra $\rho_m$ or a larger $m_{\mathrm{PBH}}$ expands faster; a PBHL with a larger $-K_c$ will keep expanding faster for ever; the motion of PBHs caused by the expansion of PBHLs does occur at speeds close to that of light. Meanwhile, we use both analytical estimates and numerical simulations to cross check the production of GWs in expanding PBHLs and find that Re$[\Psi_4^{2,0}]\propto D_{\mathrm{edge}}(\tau)'~D_{\mathrm{edge}}(\tau)''$.

Comparing with the black hole and neutron star binaries' coalescence, there is an abnormal feature that the amplitude of GWs doesn't decrease with radius but increase with radius in expanding PBHLs. Our explanation is that the final waveform of GWs at any location is the superposition of eight waveforms of GWs produced by eight adjacent PBHs in expanding PBHLs and every waveform of GWs obeys the theoretical estimation (\ref{theoretical}).

There is one caveat: the total energy density $\rho_t$ of the real Universe in matter dominated era is much smaller than our exaggerated initial values $\sim10^{-3}$. As we known, in SI units, this value corresponds to $ 6.1727\cdot 10^{17}\, \mathrm{kg/m^3}$ hence $H^2=3.45\cdot 10^{47}\mathrm{km^2/s^2/Mpc^2}$. Although we can suppress $\rho_t$ through enlarging the volume of our cubic until a reasonable $\rho_t$ compared to the real Universe in dominated matter era, we can't afford the accompanying high computational cost. Therefore, we wouldn't give the forecast whether GWs detectors can detect such signals presented in our paper or not in future.

\vspace{5mm}
\noindent {\bf Acknowledgments}
We  would  like  to  thank Xiao Guo and You-Jun Lu for their helpful discussions and advices on this paper. 
This work is partly supported by the National Natural Science Foundation of China under grant No. 11690024, the Strategic Priority Program of the Chinese Academy of Sciences (Grant No. XDB 23040100).



\end{document}